\documentclass[twocolumn]{aastex61}

\shorttitle{Correction of Near-Infrared High-Resolution Spectra for Telluric Absorption}
\shortauthors{Sameshima et al.}

\usepackage{booktabs}

\begin{document}

\title{Correction of Near-Infrared High-Resolution Spectra \\
for Telluric Absorption at 0.90--1.35 microns}

%\correspondingauthor{August Muench}
\email{sameshima@cc.kyoto-su.ac.jp}

\author{Hiroaki Sameshima}
\affiliation{Laboratory of Infrared High-resolution spectroscopy (LiH), Koyama
Astronomical Observatory, Kyoto Sangyo University, Motoyama, Kamigamo,
Kita-ku, Kyoto 603-8555, Japan}

\author{Noriyuki Matsunaga}
\affiliation{Department of Astronomy, Graduate School of Science, The
University of Tokyo, 7-3-1 Hongo, Bunkyo-ku, Tokyo 113-0033, Japan}
\affiliation{Laboratory of Infrared High-resolution spectroscopy (LiH), Koyama
Astronomical Observatory, Kyoto Sangyo University, Motoyama, Kamigamo,
Kita-ku, Kyoto 603-8555, Japan}

\author{Naoto Kobayashi}
\affiliation{Kiso Observatory, Institute of Astronomy, School of
Science, The University of Tokyo, 10762-30 Mitake, Kiso-machi, Kiso-gun,
Nagano, 397-0101, Japan}
\affiliation{Institute of Astronomy, School of Science, The University of
Tokyo, 2-21-1 Osawa, Mitaka, Tokyo 181-0015, Japan}

\author{Hideyo Kawakita}
\affiliation{Laboratory of Infrared High-resolution spectroscopy (LiH), Koyama
Astronomical Observatory, Kyoto Sangyo University, Motoyama, Kamigamo,
Kita-ku, Kyoto 603-8555, Japan}
\affiliation{Department of Physics, Faculty of Sciences, Kyoto Sangyo
University, Motoyama, Kamigamo, Kita-ku, Kyoto 603-8555, Japan}

\author{Satoshi Hamano}
\affiliation{Laboratory of Infrared High-resolution spectroscopy (LiH), Koyama
Astronomical Observatory, Kyoto Sangyo University, Motoyama, Kamigamo,
Kita-ku, Kyoto 603-8555, Japan}

\author{Yuji Ikeda}
\affiliation{Laboratory of Infrared High-resolution spectroscopy (LiH), Koyama
Astronomical Observatory, Kyoto Sangyo University, Motoyama, Kamigamo,
Kita-ku, Kyoto 603-8555, Japan}
\affiliation{Photocoding, 460-102 Iwakura-Nakamachi, Sakyo-ku, Kyoto,606-0025, Japan}

\author{Sohei Kondo}
\affiliation{Laboratory of Infrared High-resolution spectroscopy (LiH), Koyama
Astronomical Observatory, Kyoto Sangyo University, Motoyama, Kamigamo,
Kita-ku, Kyoto 603-8555, Japan}

\author{Kei Fukue}
\affiliation{Laboratory of Infrared High-resolution spectroscopy (LiH), Koyama
Astronomical Observatory, Kyoto Sangyo University, Motoyama, Kamigamo,
Kita-ku, Kyoto 603-8555, Japan}

\author{Daisuke Taniguchi}
\affiliation{Department of Astronomy, Graduate School of Science, The
University of Tokyo, 7-3-1 Hongo, Bunkyo-ku, Tokyo 113-0033, Japan}

\author{Misaki Mizumoto}
\affiliation{Department of Astronomy, Graduate School of Science, The
University of Tokyo, 7-3-1 Hongo, Bunkyo-ku, Tokyo 113-0033, Japan}
\affiliation{Institute of Space and Astronautical Science (ISAS), Japan
Aerospace Exploration Agency (JAXA), 3-1-1, Yoshinodai, Chuo-ku,
Sagamihara, Kanagawa, 252-5210, Japan}

\author{Akira Arai}
\affiliation{Laboratory of Infrared High-resolution spectroscopy (LiH), Koyama
Astronomical Observatory, Kyoto Sangyo University, Motoyama, Kamigamo,
Kita-ku, Kyoto 603-8555, Japan}

\author{Shogo Otsubo}
\affiliation{Laboratory of Infrared High-resolution spectroscopy (LiH), Koyama
Astronomical Observatory, Kyoto Sangyo University, Motoyama, Kamigamo,
Kita-ku, Kyoto 603-8555, Japan}
\affiliation{Department of Physics, Faculty of Sciences, Kyoto Sangyo
University, Motoyama, Kamigamo, Kita-ku, Kyoto 603-8555, Japan}

\author{Keiichi Takenaka}
\affiliation{Laboratory of Infrared High-resolution spectroscopy (LiH), Koyama
Astronomical Observatory, Kyoto Sangyo University, Motoyama, Kamigamo,
Kita-ku, Kyoto 603-8555, Japan}
\affiliation{Department of Physics, Faculty of Sciences, Kyoto Sangyo
University, Motoyama, Kamigamo, Kita-ku, Kyoto 603-8555, Japan}

\author{Ayaka Watase}
\affiliation{Laboratory of Infrared High-resolution spectroscopy (LiH), Koyama
Astronomical Observatory, Kyoto Sangyo University, Motoyama, Kamigamo,
Kita-ku, Kyoto 603-8555, Japan}
\affiliation{Department of Physics, Faculty of Sciences, Kyoto Sangyo
University, Motoyama, Kamigamo, Kita-ku, Kyoto 603-8555, Japan}

\author{Akira Asano}
\affiliation{Laboratory of Infrared High-resolution spectroscopy (LiH), Koyama
Astronomical Observatory, Kyoto Sangyo University, Motoyama, Kamigamo,
Kita-ku, Kyoto 603-8555, Japan}
\affiliation{Department of Physics, Faculty of Sciences, Kyoto Sangyo
University, Motoyama, Kamigamo, Kita-ku, Kyoto 603-8555, Japan}

\author{Chikako Yasui}
\affiliation{National Astronomical Observatory of Japan, 2-21-1 Osawa,
Mitaka, Tokyo 181-8588, Japan}

\author{Natsuko Izumi}
\affiliation{National Astronomical Observatory of Japan, 2-21-1 Osawa,
Mitaka, Tokyo 181-8588, Japan}

\author{Tomohiro Yoshikawa}
\affiliation{Edechs, 17-203 Iwakura-Minami-Osagi-cho, Sakyo-ku, Kyoto
606-0003, Japan}

%%%%%%%%%%%%%%%%%%%%%%%%%%%%%%%%%%%%%%%%%%
% Abstract
%%%%%%%%%%%%%%%%%%%%%%%%%%%%%%%%%%%%%%%%%%

\begin{abstract}

We report a method of correcting a near-infrared (0.90--1.35 \micron)
high-resolution ($\lambda/\Delta\lambda\sim28,000$) spectrum for
telluric absorption using the corresponding spectrum of a telluric
standard star.  The proposed method uses an A0\,V star or its analog as
a standard star from which on the order of 100 intrinsic stellar lines
are carefully removed with the help of a reference synthetic telluric
spectrum.  We find that this method can also be applied to feature-rich
objects having spectra with heavily blended intrinsic stellar and
telluric lines and present an application to a G-type giant using this
approach.  We also develop a new diagnostic method for evaluating the
accuracy of telluric correction and use it to demonstrate that our
method achieves an accuracy better than 2\% for spectral parts for which
the atmospheric transmittance is as low as $\sim$20\% if telluric
standard stars are observed under the following conditions: (1) the
difference in airmass between the target and the standard is $\lesssim
0.05$; and (2) that in time is less than 1 h.  In particular, the time
variability of water vapor has a large impact on the accuracy of
telluric correction and minimizing the difference in time from that of
the telluric standard star is important especially in near-infrared
high-resolution spectroscopic observation.

\end{abstract}

\keywords{atmospheric effects ---
 instrumentation: spectrographs --- technique: spectroscopic ---
 methods: observational --- methods: data analysis}

%%%%%%%%%%%%%%%%%%%%%%%%%%%%%%%%%%%%%%%%%%
% Introduction
%%%%%%%%%%%%%%%%%%%%%%%%%%%%%%%%%%%%%%%%%%
\section{Introduction}

Ground-based near-infrared (NIR) spectroscopy always suffers from the
absorption features of the Earth's atmosphere, which are in particular
resolved as many telluric absorption lines when the spectral resolution
is high (e.g., $R\equiv\lambda/\Delta\lambda > 10^4$).  These telluric
absorption lines must be removed to retrieve the intrinsic stellar
spectrum; however, both the strength and the line profile of telluric
absorption vary significantly with airmass and time-dependent weather
conditions, which makes removal nontrivial.

A classic but very effective method to remove telluric absorption lines
is to observe a ``telluric standard star'' and use its spectrum to
develop a correction function.  To do this, observers generally try to
make observations of the standard and target that are close in terms of
time and airmass.  If there are no intrinsic features in the spectrum of
the telluric standard star (or those features are negligible, e.g., for
a low-resolution spectrum), dividing the target spectrum by that of the
standard star should effectively remove the atmospheric telluric
absorption lines even for instruments with unusual and/or
wavelength-dependent line spread function, which is a major advantage
compared with model approaches.  In practice, however, all stars have
their own spectral features that cannot be neglected (this is
particularly true for high-resolution spectra), resulting in spectral
residuals following telluric correction.

Several methods have been proposed to overcome such problems with
telluric standard stars.  \cite{1996AJ....111..537M} proposed using
early G\,V or late F\,V stars as telluric standards.  They corrected the
stellar features of such standard stars through the fitting of a
high-resolution solar spectrum that are broadened to match the
rotational velocity induced line width of the standard star.  However,
\cite{1996ApJS..107..281H} pointed out that G\,V stars have numerous
faint metal lines that vary significantly in strength with age, chemical
composition, projected rotational velocity (hereafter referred to as $v
\sin i$), and temperature.  Thus, a perfect match between the broadened
solar spectrum and a G\,V stellar spectrum is hard to achieve, making
the method in \cite{1996AJ....111..537M} applicable only to spectra with
moderate resolutions and signal-to-noise ratios (SNRs).  As an
alternative, \cite{1996ApJS..107..281H} proposed using both an A\,V and
a G\,V star as telluric standards for $K$-band spectroscopy.  Because
the metal lines of an A\,V star are relatively infrequent and weak, they
identified removal of hydrogen lines as the main task in this correction
method.  They then created a telluric spectrum from the observed G\,V
star in the same manner as \cite{1996AJ....111..537M} and removed it
from the Br\,$\gamma$ region of the A\,V star.  The resulting cleaned
Br\,$\gamma$ feature was then fitted and removed to produce the final
A\,V stellar spectrum representing the pure telluric absorption.  This
method combines the advantages of methods using A\,V stars and those
using G\,V stars but is relatively expensive with respect to observation
time.  \cite{2003PASP..115..389V} improved the method of
\cite{1996AJ....111..537M} by focusing on the use of A0\,V stars as
telluric standards for which the stellar features were corrected by
fitting a high-resolution model spectrum of Vega.  Except for some
high-order Paschen lines, their method does a good job of removing A0\,V
hydrogen lines and can successfully remove telluric absorption lines in
low-to-medium resolution spectra ($R \sim200$--2500) in the $J$, $H$,
and $K$ bands.  It is worth noting that, despite the high number of
studies on low- and medium-resolution spectra, there have been no
systematic efforts to establish methods for using telluric standard
stars to remove the telluric features of NIR high-resolution ($R \gtrsim
10,000$) spectra.

In recent years, much effort has been put into developing methods for
using synthetic telluric spectra to correct telluric absorption in
observed spectra (e.g., \citealt{2010A&A...524A..11S,
2014MNRAS.439..387C, 2014A&A...564A..46B, 2014AJ....148...53G,
2016A&A...585A.113R}).  Such model approaches have the significant
advantages of avoiding intrinsic-line problems associated with telluric
standard stars, being free of noise due to observation that cannot be
avoided in classic approaches using telluric standard stars, and saving
precious telescope time by not requiring observations of telluric
standard stars.  One of the most powerful of these tools is {\tt
molecfit} (\citealt{2015A&A...576A..77S};
\citealt{2015A&A...576A..78K}), which incorporates the radiative
transfer code LBLRTM (\citealt{2005JQSRT..91..233C}), the line database
HITRAN (\citealt{2009JQSRT.110..533R}), a combination of meteorological
data from various sources, and a model of the line spread function.
\cite{2015A&A...576A..77S} reported that the accuracy of {\tt molecfit},
as measured by the standard deviation of the residuals after correction
of unsaturated telluric lines, is frequently better than 2\% of the
continuum, which is comparable to the typical accuracy achieved using a
telluric standard star.  These model approaches, however, have
fundamental limitations caused by missed or uncertain information
regarding lines in molecular databases, imperfect modeling of line
profiles especially for instruments with an unusual and/or
wavelength-dependent line spread function, limited knowledge of
atmospheric conditions, and treatment of wind.  Furthermore, as
\cite{2010A&A...524A..11S} mentioned, it is not an easy task to find a
synthetic telluric spectrum that gives the best match with the telluric
absorption in the spectrum of a feature-rich object such as a late-type
star whose intrinsic stellar lines are severely blended with telluric
lines.  Therefore, application of classical methods using telluric
standard stars to NIR high-resolution spectra for correcting telluric
absorption are still worth investigating, especially for high-SNR
spectra.

Here, we present a method for telluric correction developed for
high-resolution and high-SNR NIR spectra obtained via the WINERED
echelle spectrograph (\citealt{2016SPIE.9908E..5ZI}).  WINERED uses a
1.7 \micron-cutoff $2048\times2048$ HAWAII--2RG infrared array,
simultaneously covering the wavelength range 0.90--1.35 \micron\ without
any instrumental gap.  The free spectral range of each spectral order is
summarized in Table \ref{tab:order}, where the main atmospheric
absorbers in the wavelength range are also given.  The slit width is 100
\micron\footnote{With WINERED attached to the 1.3-m Araki telescope at
the Koyama Astronomical Observatory in Kyoto, Japan, this slit width
corresponds to 1\arcsec.6 on the sky and the pixel scale is 0\arcsec.8
pixel$^{-1}$.}, resulting in a spectral resolution of $R \sim 28,000$.
Scientific results based on WINERED observations have been previously
reported (e.g., \citealt{2015ApJ...800..137H,2016ApJ...821...42H};
\citealt{2018MNRAS.473.4993T}).  The method proposed in this paper uses
A0\,V stars, or their analogs, as telluric standard stars because their
spectra are relatively featureless and, unlike OB stars, they do not
have strong helium lines or emission lines associated with surrounding
gas.  Using synthetic telluric spectra created with {\tt molecfit} as a
reference, the proposed method identifies intrinsic stellar lines from
A0\,V stars and removes them from observed spectra.  As an example,
\cite{2018MNRAS.473.4993T} used our method to measure the line-depth
ratios of late-type giant spectra obtained using WINERED.

This paper is organized as follows.  The mathematical concepts
underlying our method are given in \S2, while details of the practical
procedures it employs are given in \S3.  Examples of telluric absorption
correction for a G-type star and for OB stars using our method are
presented in \S4.  In \S5, the quality of the proposed method is
measured through comparison with a method for simple spectral division
by early-type stars and via a newly developed diagnostic method, and its
dependence on observing conditions is discussed.  Finally, a brief
summary is given in \S6.  Throughout this paper, we use air wavelengths,
rather than vacuum wavelengths, unless otherwise noted.

\begin{deluxetable}{lcc}
\tablecaption{Spectral orders of WINERED \label{tab:order}}
\tablehead{
\colhead{Order} & \colhead{Free Spectral Range (\AA)} & \colhead{Absorbers\tablenotemark{a}}
 }
 \startdata
 61 &  9126--9275  & H$_2$O \\
 60 &  9275--9432  & H$_2$O \\
 59 &  9432--9592  & H$_2$O \\
 58 &  9592--9759  & H$_2$O \\
 57 &  9759--9933  & H$_2$O \\
 56 &  9933--10113 & H$_2$O \\
 55 & 10113--10296 & H$_2$O \\
 54 & 10296--10489 & H$_2$O \tablenotemark{b} \\
 53 & 10489--10688 & H$_2$O \tablenotemark{b} \\
 52 & 10688--10894 & H$_2$O \\
 51 & 10894--11108 & H$_2$O \\
 50 & 11108--11337 & H$_2$O \\
 49 & 11337--11565 & H$_2$O \\
 48 & 11565--11810 & H$_2$O \\
 47 & 11810--12063 & H$_2$O, CO$_2$ \\
 46 & 12063--12327 & H$_2$O, CO$_2$ \\
 45 & 12327--12603 & O$_2$, H$_2$O \\
 44 & 12603--12895 & O$_2$, H$_2$O \\
 43 & 12895--13190 & H$_2$O \\
 42 & 13190--13509 & H$_2$O \\
 \enddata
 \tablenotetext{a}{Only predominant contributors are listed.}
 \tablenotetext{b}{Telluric absorption is nearly negligible in this order.}
\end{deluxetable}

%%%%%%%%%%%%%%%%%%%%%%%%%%%%%%%%%%%%%%%%%%
% Methodology
%%%%%%%%%%%%%%%%%%%%%%%%%%%%%%%%%%%%%%%%%%

\section{Methodology} \label{sec:methodology}

Following \cite{2003PASP..115..389V}, we write the observed spectrum
$O_{\mathrm{std}}(\lambda)$ of a telluric standard star as
\begin{equation}
 O_{\mathrm{std}}(\lambda) = [I_{\mathrm{std}}(\lambda) \cdot
  T(\lambda; A_{\mathrm{std}})] * P(\lambda) \cdot Q(\lambda),
\end{equation}
where $I_{\mathrm{std}}(\lambda)$ is the intrinsic spectrum of the
telluric standard star, $T(\lambda; A_{\mathrm{std}})$ is the telluric
absorption spectrum at the airmass of $A_{\mathrm{std}}$, $P(\lambda)$
is the instrumental profile, $Q(\lambda)$ is the instrumental
throughput, and an asterisk denotes convolution.  Here, we assume that
$Q(\lambda)$ is a featureless smooth function and is not affected by
convolution with $P(\lambda)$.

The purpose of this telluric correction is to extract, as accurately as
possible, information on $T(\lambda; A_{\mathrm{std}})$ from
$O_{\mathrm{std}}(\lambda)$.  To this end, the observed spectrum of the
telluric standard star is analyzed using the following steps.

First, the observed spectrum is normalized so that the continuum is
unity.  This yields 
\begin{equation}
 O_{\mathrm{std}}^\prime(\lambda) = [I_{\mathrm{std}}^\prime(\lambda)
  \cdot T(\lambda; A_{\mathrm{std}})] * P(\lambda),
\end{equation}
where the prime symbol denotes continuum normalization.  As was done by
\cite{2003PASP..115..389V}, this equation is often approximated as the
multiplicative product of a smoothed stellar spectrum and a smoothed
telluric spectrum:
\begin{equation}
 O_{\mathrm{std}}^\prime(\lambda) \sim [I_{\mathrm{std}}^\prime(\lambda)
  * P(\lambda)] \cdot [T(\lambda; A_{\mathrm{std}}) * P(\lambda)]. \label{eq:approx}
\end{equation}
Note, however, that this approximation is only applicable if the stellar
lines are spectrally resolved; see the appendix for more discussion on
this.

Next, a synthetic telluric absorption spectrum convolved with an
instrumental profile $[T(\lambda; A_{\mathrm{std}}) *
P(\lambda)]_{\mathrm{model}}$ is created using {\tt molecfit}
(\citealt{2015A&A...576A..77S}; \citealt{2015A&A...576A..78K}).
Dividing $O_{\mathrm{std}}^\prime(\lambda)$ by this synthetic spectrum,
we have
\begin{equation}
 R(\lambda) = [I_{\mathrm{std}}^\prime(\lambda) * P(\lambda)] \cdot \Delta(\lambda),
\end{equation}
where $\Delta(\lambda) \equiv [T(\lambda; A_{\mathrm{std}}) *
P(\lambda)] / [T(\lambda; A_{\mathrm{std}}) *
P(\lambda)]_{\mathrm{model}}$ denotes the difference between the model
and observation.  $\Delta(\lambda)$ is nearly flat at unity, but
residuals do exist; it is probable that this is largely caused by
inaccuracies in modeling the instrumental profile $P(\lambda)$ of
WINERED.  Despite the presence of such noise, $R(\lambda)$ reflects
$[I_{\mathrm{std}}^\prime(\lambda) * P(\lambda)]$ sufficiently well to
enable us to identify the intrinsic lines of telluric standard stars.

Each intrinsic stellar line present in $R(\lambda)$ is then fitted with
multiple Gaussian curves.  Combining these fitted intrinsic lines
produces the spectrum of the telluric standard star,
$[I_{\mathrm{std}}^\prime(\lambda) * P(\lambda)]_{\mathrm{fit}}$.
Dividing $O_{\mathrm{std}}^\prime(\lambda)$ by this spectrum, we obtain 
\begin{equation}
 S(\lambda; A_{\mathrm{std}}) = [T(\lambda; A_{\mathrm{std}}) * P(\lambda)] \cdot \sigma(\lambda), 
\end{equation}
where $\sigma(\lambda) = [I_{\mathrm{std}}^\prime(\lambda) * P(\lambda)]
/ [I_{\mathrm{std}}^\prime(\lambda) * P(\lambda)]_{\mathrm{fit}}$
denotes the difference between the intrinsic and fitted spectra.
$S(\lambda; A_{\mathrm{std}})$ now represents the telluric absorption
imprinted in the spectrum of the telluric standard star.

To apply $S(\lambda; A_{\mathrm{std}})$ to the spectrum of a target
object, the difference in airmass between the target and the telluric
standard star must be corrected.  Following Beer's law
(\citealt{1852AnP...162...78B}), the corrected telluric absorption
spectrum $S(\lambda; A_{\mathrm{tgt}})$ is written as
\begin{equation}
 S(\lambda; A_{\mathrm{tgt}}) = S(\lambda -
  \alpha; A_{\mathrm{std}})^{\beta
  \frac{A_{\mathrm{tgt}}}{A_{\mathrm{std}}}}, \label{eq:beer}
\end{equation}
where $\alpha$ is the wavelength shift between the target and the
standard and $\beta$ represents effects other than the airmass ratio
(such as the time variability of the absorber's number density).  The
values of $\alpha$ and $\beta$ can be determined using the IRAF task
{\tt telluric} by minimizing the root-mean-square values of the
corrected spectrum in the selected wavelength region.

Finally, dividing the observed spectrum of the target
$O(\lambda)_{\mathrm{tgt}}$ by $S(\lambda; A_{\mathrm{tgt}})$, we obtain
a telluric-corrected spectrum:
\begin{equation}
 G(\lambda) = [I_{\mathrm{tgt}}(\lambda) * P(\lambda)] \cdot Q(\lambda)
  \cdot \sigma(\lambda),
  \label{eq:target_corrected}
\end{equation}
from which the telluric absorption term $T(\lambda; A_{\mathrm{tgt}})$
has been removed.

%%%%%%%%%%%%%%%%%%%%%%%%%%%%%%%%%%%%%%%%%%
% Detailed procedures
%%%%%%%%%%%%%%%%%%%%%%%%%%%%%%%%%%%%%%%%%%

\section{Detailed procedures} \label{sec:details}

\subsection{Continuum normalization}

\begin{figure}[t!]
 \epsscale{1.1}
 \plotone{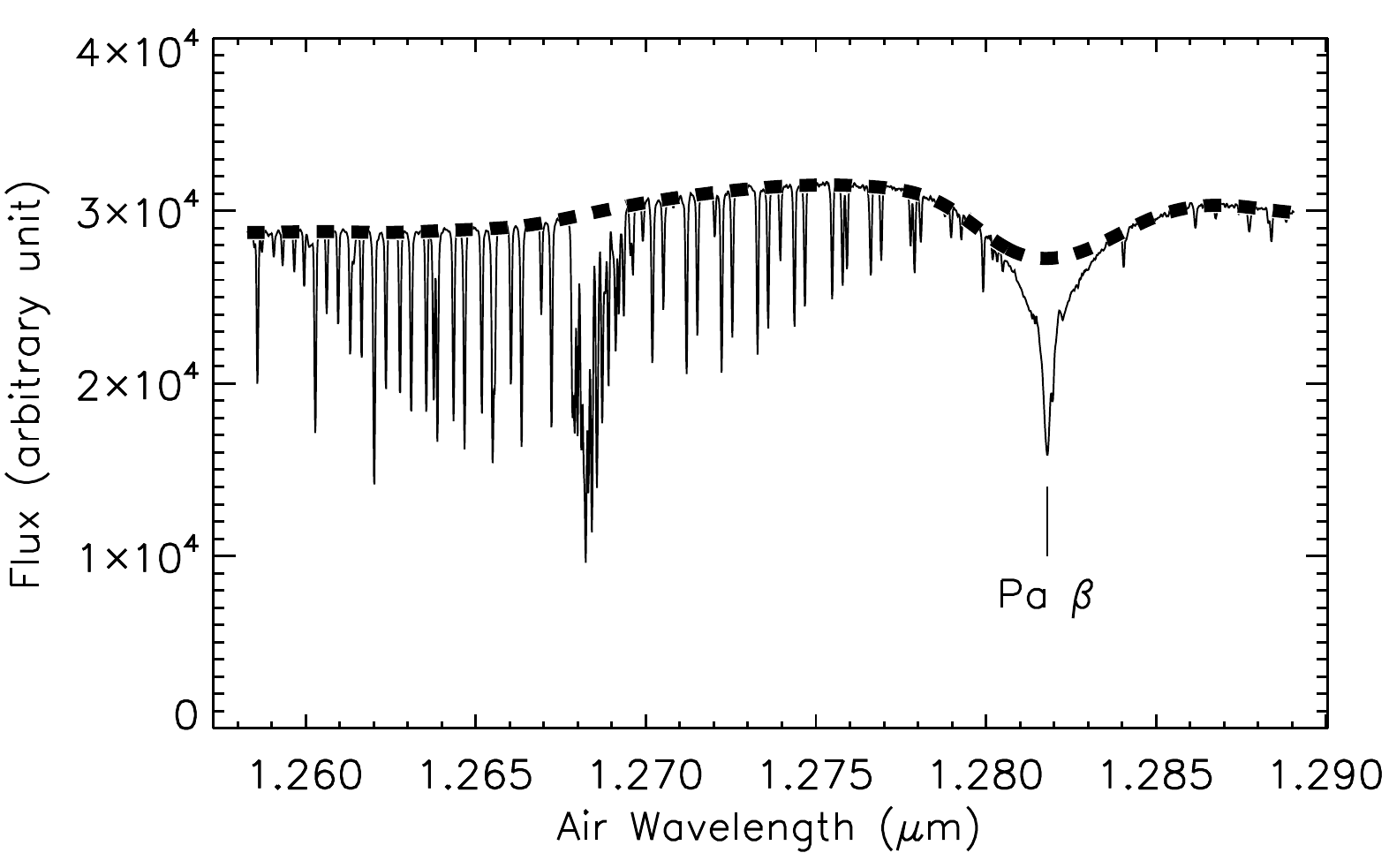}
 \caption{Continuum normalization of an A0\,V star (order 44).  The
 solid line indicates the observed spectrum, while the dashed line
 indicates the fitted continuum spectrum determined using the IRAF task
 {\tt continuum}.  Note that the majority of the absorption lines seen
 in this figure are from telluric absorption (mainly caused by O$_2$
 molecules in this wavelength region).  As discussed in the text, the
 continuum obtained here is not suitable for studying the Pa\,$\beta$
 line but works sufficiently for our purpose of using this spectrum for
 telluric correction.  } \label{fig:21Lyn_continuum}
\end{figure}

To perform continuum normalization of the observed spectrum of a
telluric standard star, we use the IRAF task {\tt continuum}.
Generally, a cubic spline curve is used to fit the continuum, but
Legendre polynomials can also be used depending on the situation.  An
example of continuum normalization of an A0\,V star is shown in Figure
\ref{fig:21Lyn_continuum}, from which it is seen that the {\tt
continuum} task can estimate the continuum well even in a relatively
crowded telluric absorption area although the wing part of Pa\,$\beta$
is partly confused with the continuum.  Such confusion partially wipes
out the broad wing of hydrogen lines and narrows the lines remaining in
the normalized spectrum relative to those in the real spectrum.
However, we have found that in our method it is easier to transform
normalized spectra such as these without very broad lines into telluric
spectra than it is to transform spectra with the real line profiles of
broad hydrogen lines.  This is because, in the subsequent step, {\tt
molecfit} does not have to account for the contribution from the widely
spread wing parts of hydrogen lines when fitting a synthetic telluric
spectrum around hydrogen lines.  Note that we are not interested in the
true profiles of the hydrogen lines during the telluric correction.

Although the {\tt continuum} task skillfully estimates the continua of
observed spectra for most orders, it often fails at around 9,300--9,600
\AA, where telluric absorption is very severe.  In such cases, the
observed spectrum is first processed using {\tt molecfit} for rough
removal of telluric absorption lines and the {\tt continuum} task is
then performed to estimate the continuum level.  This additional
procedure often improves the results of the continuum-normalized
spectrum.

\subsection{Making synthetic telluric spectra} \label{sec:molecfit}

The next step is to use {\tt molecfit} to create a synthetic telluric
spectrum $[T(\lambda; A_{\mathrm{std}}) * P(\lambda)]_{\mathrm{model}}$
using the normalized spectrum of the telluric standard star.  It should
be noted that in our method the created synthetic spectrum is used only
as a reference for identifying spectral features other than telluric
absorption on the spectrum of a telluric standard star.  At this step,
it is unnecessary to fine-tune the {\tt molecfit} parameters to remove
the telluric lines perfectly, but finding a reasonably good solution is
useful in making the following steps simple and efficient.  Some key
points regarding tuning the settings of {\tt molecfit} to achieve such
accuracy are described below, but readers are also referred to
\cite{2015A&A...576A..77S} and to the {\tt molecfit} manual for more
details on its functions.

We run {\tt molecfit} for each spectral order independently to create
synthetic telluric spectra.  The instrumental profile $P(\lambda)$ is
determined through a fitting in which $P(\lambda)$ is modeled as a
combination of Gaussian and Lorentzian curves.  Only the molecules
listed in the last column in Table \ref{tab:order} are included in the
calculation and fitted for each spectral order, and the other molecules
are switched off because they have negligible contribution on results.
We have found that the initial value of the H$_2$O scaling
factor\footnote{The relevant parameter in {\tt molecfit} is called {\tt
RELCOL\_H2O}.  It scales the atmospheric profile, i.e., vertical
distribution, of H$_2$O molecules while keeping the profile shape, which
is independently determined by the given relative humidity, unchanged.}
often has significant impacts on the fit, and a careful tuning of this
parameter is one of the most important factors in achieving high
accuracy.  In addition, masking spectral features other than telluric
absorption is also important.  The wavelength ranges holding the
intrinsic A0\,V star absorption lines as predicted by a model spectrum,
created by using {\tt SPTOOL} (Takeda, private communication) to
implement {\tt ATLAS9} programs (\citealt{1993sssp.book.....K}), are
masked.  Other spectral features including artifacts are also checked by
eye and masked if necessary.

\begin{figure}[t!]
 \epsscale{1.1}
 \plotone{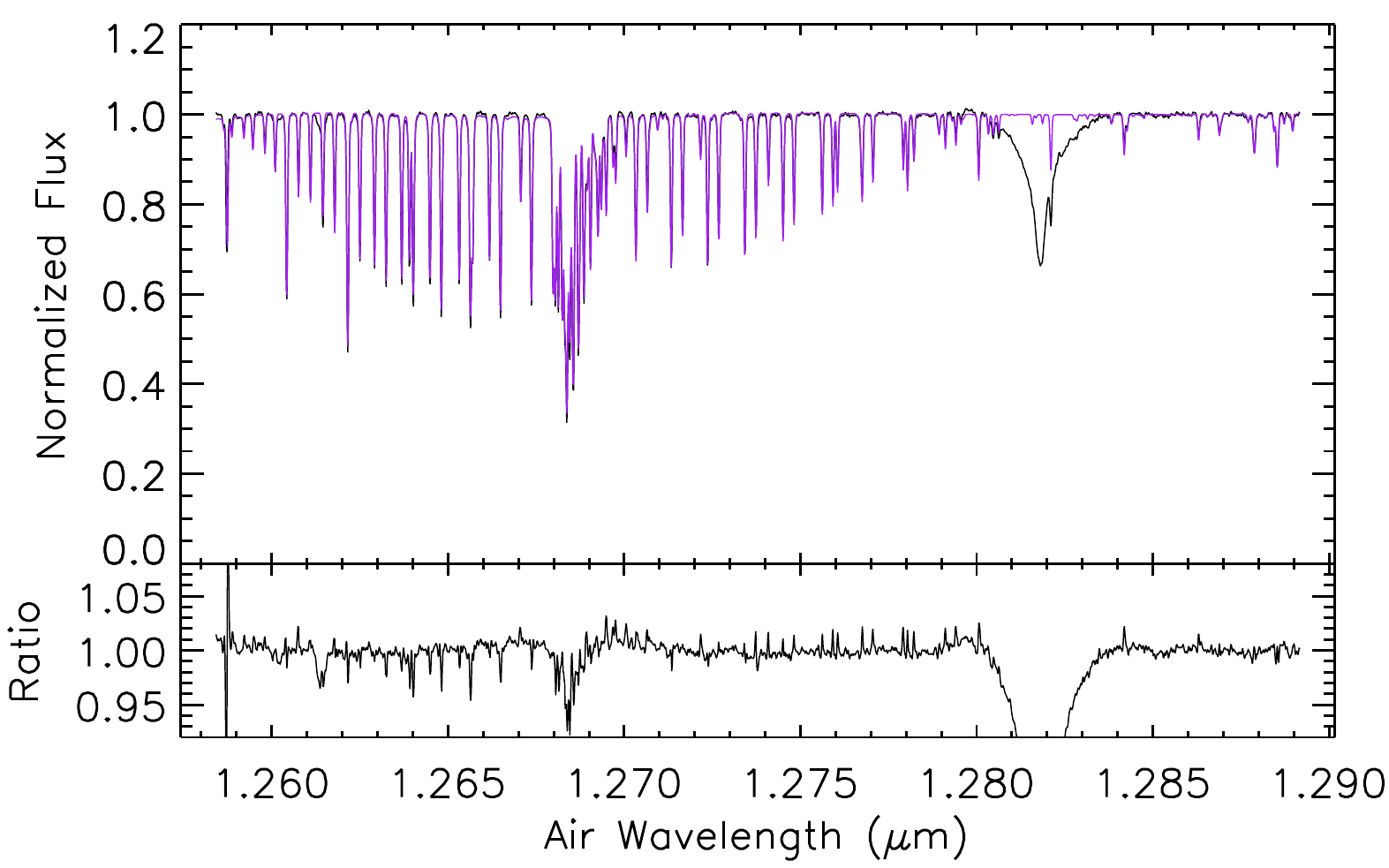}
 \caption{{\it Top:} Comparison between a synthetic telluric spectrum
 created using {\tt molecfit} and an observed spectrum.  The black line
 indicates the normalized telluric standard spectrum
 $O^\prime_{\mathrm{std}}(\lambda)$, while the purple line indicates the
 synthetic telluric spectrum $[T(\lambda; A_{\mathrm{std}}) *
 P(\lambda)]_{\mathrm{model}}$.  {\it Bottom:} $R(\lambda)$, the ratio
 of the observed spectrum to the synthetic telluric spectrum.  }
 \label{fig:molecfit_m44}
\end{figure}

An example of a synthetic telluric spectrum $[T(\lambda;
A_{\mathrm{std}}) * P(\lambda)]_{\mathrm{model}}$ created by {\tt
molecfit} for a specific spectral order is shown in Figure
\ref{fig:molecfit_m44}.  From this figure, we observe that the
difference between observation and model, or $\Delta(\lambda)$, has
uncertainties at a few percent level, which is rather large in
comparison to the uncertainties obtained by classical methods using
telluric standard spectra.  This result is likely caused by poor
accuracy of modeling of the instrumental profile $P(\lambda)$ and the
slight wavelength uncertainties of the WINERED spectra.  Further fine
tuning of {\tt molecfit} parameters can improve these results, although
noisy features of a few percent were persistent during our analysis.  As
mentioned above, the purpose of using {\tt molecfit} at this step is not
to completely remove the telluric lines by itself.

\subsection{Fitting intrinsic stellar lines}

Intrinsic stellar lines of an A0\,V star present in the ratio spectrum
$R(\lambda)$ are manually fitted and subsequently removed by visual
inspection with the help of an A0\,V star model spectrum created by {\tt
SPTOOL}\footnote{We use {\tt SPTOOL} only to make template spectra of an
A0 V star for reference and never use it for line fittings. Once the
stellar lines of a standard star were identified, they were empirically
fitted with multiple Gaussian curves.  Therefore the fittings can be
performed even if non-LTE effects affect the line shape/depth of the
standard star.}.  Multiple Gaussian curves are used as a fitting
function, and fittings are performed by minimizing chi-squared values
using the IDL script {\tt mpfit.pro} (\citealt{2009ASPC..411..251M}),
which uses the Levenberg-Marquardt method.  It should be noted that, in
addition to intrinsic stellar lines, features other than telluric
absorption such as residuals of sky subtraction and
flat-fielding-related incompleteness appear in the ratio spectrum.
These are carefully removed at this stage by fitting multiple Gaussian
curves.  Generally speaking, identification and fitting are relatively
easy for narrow and deep absorption lines; for this reason, A0\,V stars
with $v \sin i \lesssim 50$ km s$^{-1}$ are generally selected as
telluric standard stars in the proposed method.

\begin{figure}[t!]
 \epsscale{1.1}
 \plotone{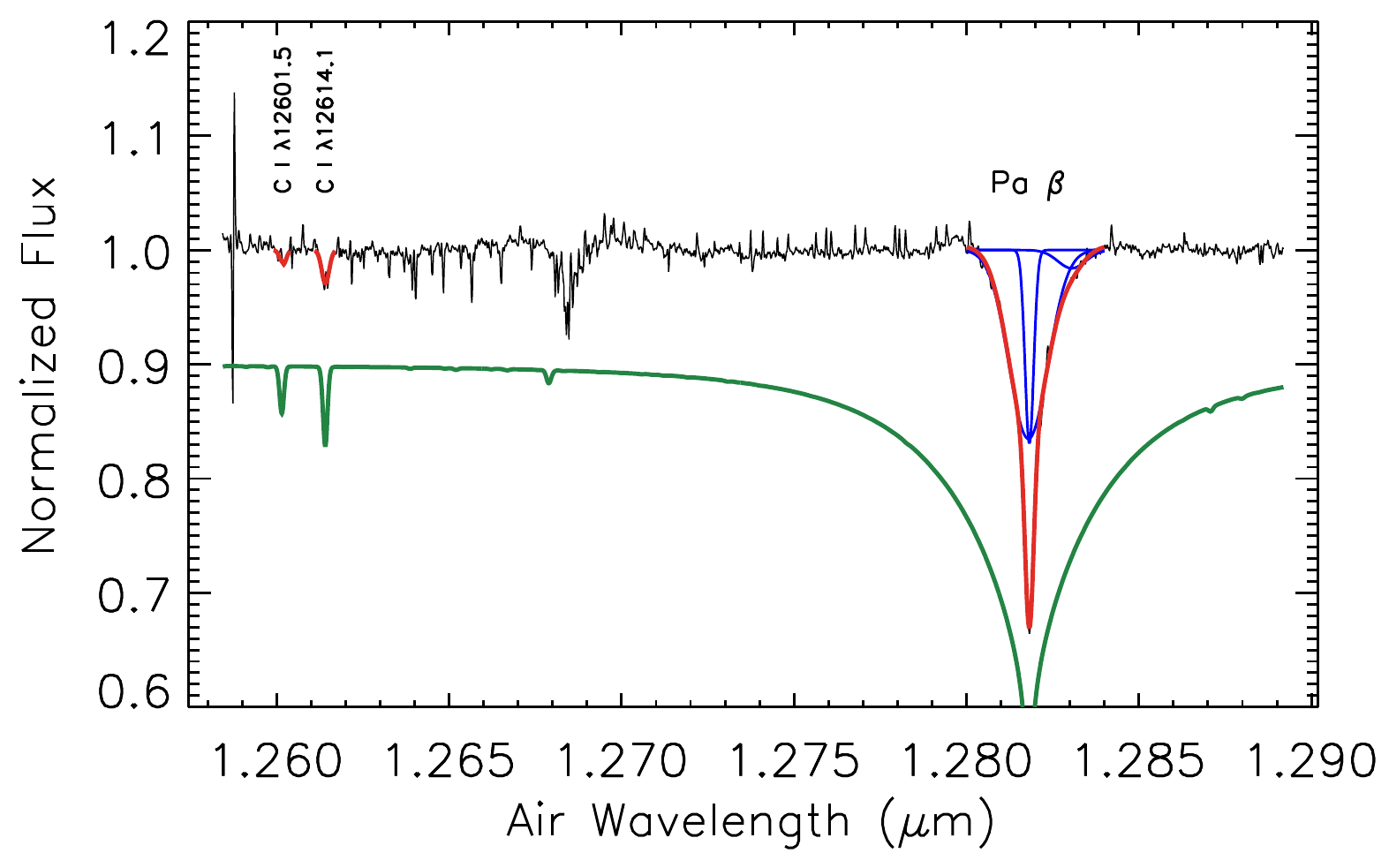}
 \caption{Schematic illustrating how the intrinsic stellar lines of an
 A0\,V star are identified and fitted.  The black line indicates the
 ratio spectrum $R(\lambda)$, i.e., the observed spectrum of a telluric
 standard star divided by a synthetic telluric spectrum.  The green line
 indicates the model spectrum of an A0\,V star created using {\tt
 SPTOOL} and shifted in vertical direction for visual clarity.
 The fitted intrinsic stellar lines, indicated by the red curves, are
 obtained by combining up to three Gaussian functions, each of which is
 indicated by a blue curve.  } \label{fig:fitting_intrinsic_lines}
\end{figure}

An example of fitting of the intrinsic stellar lines of an A0\,V star is
shown in Figure \ref{fig:fitting_intrinsic_lines}.  In this case, the
Pa\,$\beta$ profile is asymmetric owing to an error in the continuum
normalization.  Such spectra are unsuitable for studying the Pa\,$\beta$
line itself, but a (skewed) profile could be reproduced by combining
three Gaussian curves.  Moreover, we could also identify two weak metal
lines---\ion{C}{1} $\lambda\lambda 12601, 12614$---in this spectral
order.  Although these weak lines were somewhat blended with the noisy
features due to the imperfect matching of synthetic telluric spectra in
the previous step, we could fit a single Gaussian curve by masking a
couple of pixels showing significant deviations.  By contrast,
\ion{C}{1} $\lambda12679$ could not be fitted because it appeared to be
very weak and seriously blended with deep telluric absorption.  Over the
entire wavelength range of WINERED, i.e., from 0.90--1.35 \micron, we
identified more than 100 metal lines, with depths as shallow as
$\sim$1\%.  This result directly suggests the importance of removing
intrinsic stellar lines in NIR high-resolution spectroscopy, even if a
featureless A-type star is used as a telluric standard star.

%%%%%%%%%%%%%%%%%%%%%%%%%%%%%%%%%%%%%%%%%%
% Results
%%%%%%%%%%%%%%%%%%%%%%%%%%%%%%%%%%%%%%%%%%

\section{Results} \label{sec:results}

\begin{deluxetable*}{lccccccccc}
\tablecaption{Observing Log \label{tab:obslog}}
\tablehead{
\colhead{Name} & \colhead{Spectral Type\tablenotemark{a}} & \colhead{UT
 Date} & \colhead{UT Start} & \colhead{Airmass} & \colhead{Seeing} &
 \colhead{$T_\mathrm{air}$} & \colhead{RH\tablenotemark{b}} & \colhead{Exposure} &
 \colhead{SNR\tablenotemark{d}} \\
 \colhead{} & \colhead{} & \colhead{} & \colhead{} & \colhead{} &
 \colhead{(arcsec)} &
 \colhead{($^{\circ}$C)} & \colhead{(\%)} &
 \colhead{(sec)} & \colhead{}
 }
 \startdata
 HD 2905           & B1\,Ia          & 2014 Jan 22 & 09:55:26 & 1.24--1.27 & 4.3 & $+0.9$ & 78 & 720  & 830  \\
 HD 23180          & B1\,III         & 2014 Jan 22 & 12:20:48 & 1.08--1.08 & 4.0 & $-0.1$ & 84 & 400  & 600  \\
 7 Cam             & A1\,V           & 2014 Jan 22 & 12:35:16 & 1.07--1.08 & 3.8 & $-0.2$ & 84 & 960  & 750  \\
 HD 30614          & O9\,Ia          & 2014 Jan 22 & 13:02:26 & 1.19--1.21 & 3.4 & $-0.2$ & 85 & 960  & 810  \\
 HD 37022          & O7\,V           & 2014 Jan 22 & 13:34:36 & 1.38--1.42 & 3.5 & $+0.2$ & 84 & 960  & 700  \\
 HD 25204          & B4\,IV          & 2014 Jan 23 & 10:57:21 & 1.08--1.09 & 3.9 & $+1.6$ & 82 & 720  & 1040 \\
 HD 36371          & B4\,Ib          & 2014 Jan 23 & 11:18:31 & 1.03--1.02 & 3.9 & $+1.3$ & 83 & 720  & 800  \\
 HD 36822          & B0\,III         & 2014 Jan 23 & 12:15:43 & 1.11--1.11 & 4.0 & $+0.8$ & 85 & 1200 & 740  \\
 HD 37742          & O9.2\,Ib        & 2014 Jan 23 & 12:46:56 & 1.26--1.26 & 4.4 & $+0.3$ & 86 & 240  & 1160 \\
 HD 37043          & O9\,III         & 2014 Jan 23 & 13:08:07 & 1.35--1.40 & 4.1 & $+0.4$ & 85 & 600  & 890  \\
 HD 41117          & B2\,Ia          & 2014 Jan 23 & 13:49:34 & 1.06--1.07 & 3.7 & $+0.2$ & 85 & 720  & 760  \\
 HD 36486          & B0\,III + O9\,V & 2014 Jan 23 & 14:08:37 & 1.38--1.40 & 3.9 & $+0.1$ & 85 & 240  & 810  \\
 HD 43384          & B3\,Iab         & 2014 Jan 23 & 14:26:38 & 1.07--1.10 & 3.7 & $+0.0$ & 86 & 1440 & 580  \\
 21 Lyn            & A0.5\,V         & 2014 Jan 23 & 15:00:30 & 1.04--1.06 & 3.5 & $-0.1$ & 85 & 1200 & 830  \\
 $\varepsilon$ Leo & G1\,IIIa        & 2014 Jan 23 & 16:44:32 & 1.02--1.03 & 3.3 & $-0.8$ & 85 & 110  & 1010 \\
 \enddata
 \tablenotetext{a}{SIMBAD spectral type.}
 \tablenotetext{b}{Relative humidity.}
 \tablenotetext{c}{SNR is measured from the standard deviation of the 
 continuum level at spectral order 54, where telluric absorption
 is nearly negligible.}
\end{deluxetable*}

\begin{figure*}[t!]
 \epsscale{1.1}
 \plotone{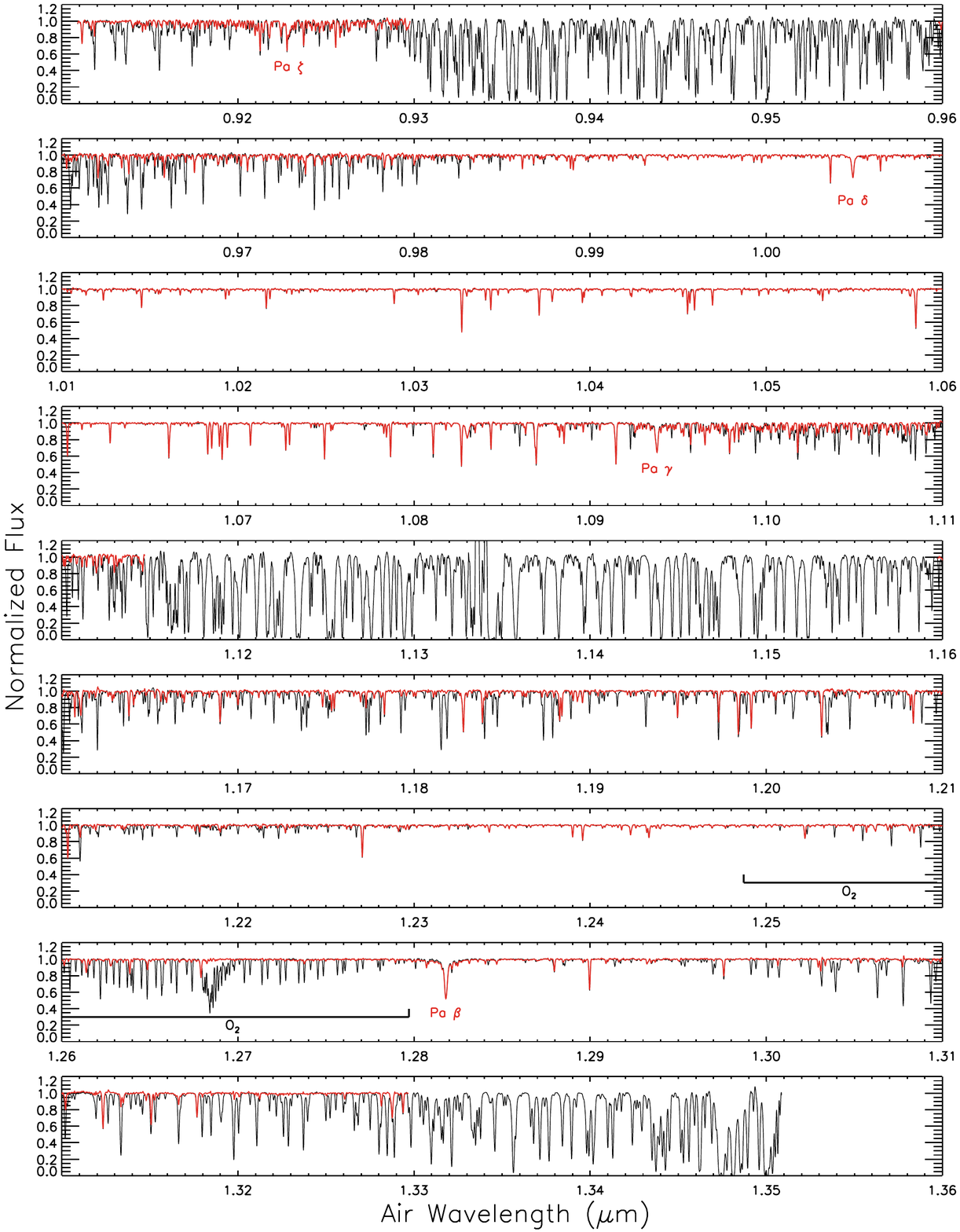}
 \caption{WINERED spectrum of $\varepsilon$ Leo (G1\,IIIa) before
 (black) and after (red) telluric correction.  The wavelength
 ranges 0.930--0.960, 1.115--1.160, and $>$1.330 \micron,
 are not treated with telluric correction owing to the heavy telluric
 absorption in these ranges.}
 \label{fig:epsLeo_all_order}
\end{figure*}

Here, we show some examples of applying the method described in
\S\S\ref{sec:methodology}--\ref{sec:details} to spectra obtained using
WINERED.  The observing log of all target objects and telluric standard
stars is summarized in Table \ref{tab:obslog}.  In the following
analysis, 7 Cam (A1\,V) and 21 Lyn (A0.5\,V) were used as telluric
standard stars for target objects observed on Jan 22 and 23, 2014,
respectively.  For these observations, WINERED was attached to the 1.3-m
Araki telescope at the Koyama Astronomical Observatory.  The air
temperature outside the dome changed from $+1.5$ to $-1 ^{\circ}$C and
the relative humidity changed from $76$ to $86$\% during the nights,
which are typical conditions of the winter season of the site.  Seeing
ranged from 3\arcsec.3 to 4\arcsec.4, which were slightly worse than the
typical value of the site ($\sim$3\arcsec.0).

Figure \ref{fig:epsLeo_all_order} compares the spectra before and after
telluric correction obtained from the G1\,III giant $\varepsilon$ Leo
over the entire wavelength range covered by WINERED.  Because telluric
absorption was too heavy in the wavelength ranges 0.930--0.960,
1.115--1.160, and $>$1.330 \micron, which correspond to the gaps between
the adjacent photometric bands ($z^\prime, Y$ and $J$), we did not
perform any telluric correction for these ranges.  Apart from these
extreme regions, our method effectively removed telluric absorption
lines even in, e.g., the strong O$_2$ band (1.25--1.28 \micron) and
successfully reproduced the intrinsic spectrum of $\varepsilon$ Leo.

\begin{figure*}[t!]
 \epsscale{1.1}
 \plotone{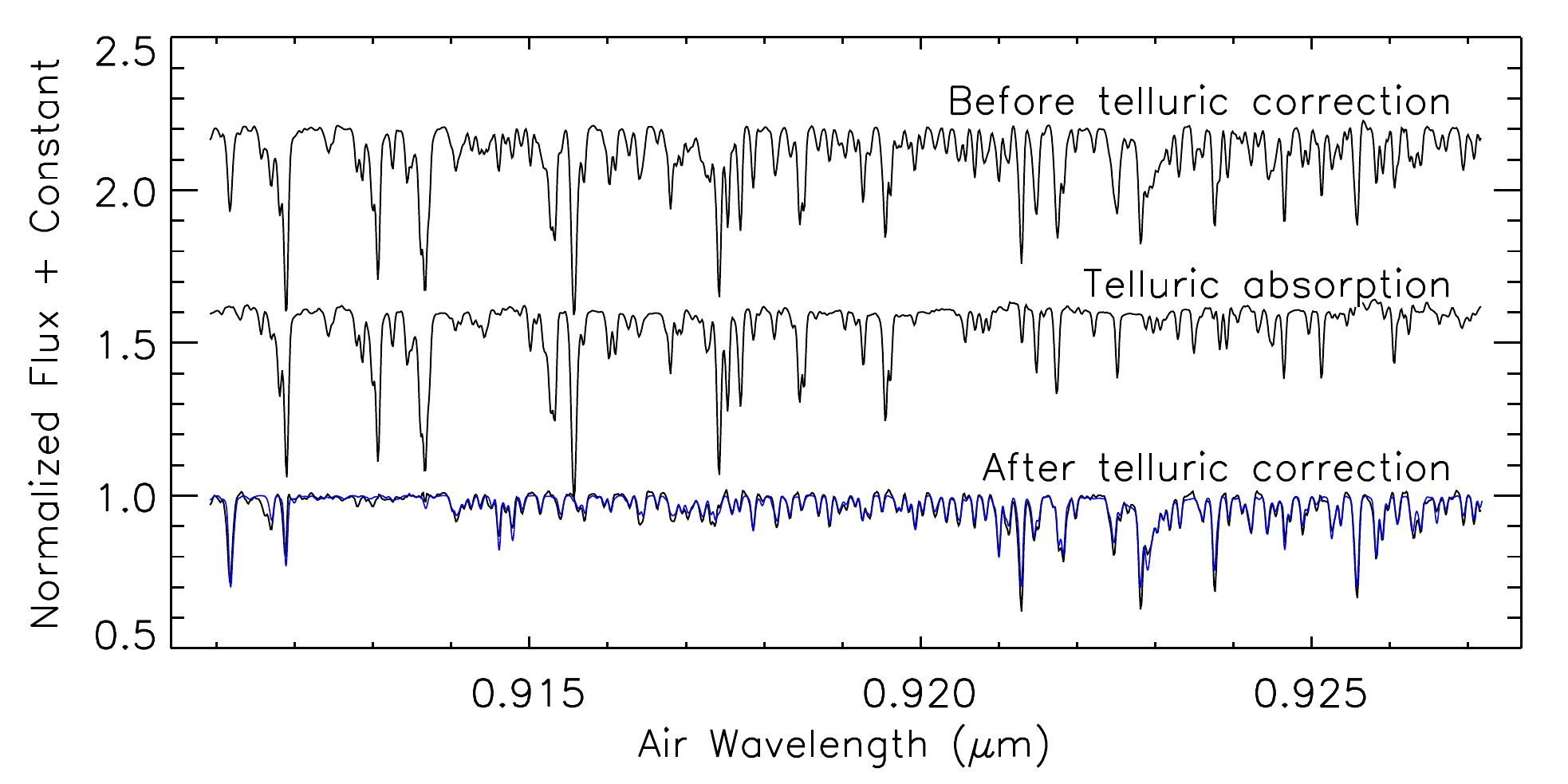}
 \caption{Telluric correction of $\varepsilon$ Leo (G1\,IIIa) for 
 spectral order 61.  The spectra are shown from top to bottom as
 follows: the raw spectrum of $\varepsilon$ Leo, the telluric absorption
 spectrum $S(\lambda; A_{\mathrm{tgt}})$ created from 21 Lyn (A0.5\,V),
 and the telluric-corrected spectrum of $\varepsilon$ Leo.  A model
 spectrum of $\varepsilon$ Leo created by {\tt SPTOOL} is indicated in
 blue.  For visual clarity, constants are added to the top and
 middle spectra.}  \label{fig:epsLeo_m61}
\end{figure*}

Figure \ref{fig:epsLeo_m61} plots the spectrum of $\varepsilon$ Leo in
the 61st order before and after telluric correction along with the
telluric absorption spectrum.  In the original spectrum, stellar lines
are heavily contaminated by telluric absorption lines, which prevented
quantitative investigations of intrinsic features.  By contrast, the
spectrum obtained after telluric correction is in close agreement with
the model stellar spectrum created by {\tt SPTOOL}\footnote{The
following model atmospheric parameters for $\varepsilon$ Leo were
adopted from \cite{2011A&A...531A.165P}: effective temperature
$T_\mathrm{eff}=5,398$ K; logarithm of the surface gravity $\log
g=2.02$; and metallicity [Fe/H]=$-0.06$.}, demonstrating the
applicability of our method to even feature-rich objects.  The
sophisticated telluric correction we obtained enabled us to identify a
large number of weak metal and CN molecular lines for $\varepsilon$ Leo,
which we will compile as a line list and report in a forthcoming paper
(Ikeda et al., in preparation).

\begin{figure*}[t!]
 \epsscale{1.1}
 \plotone{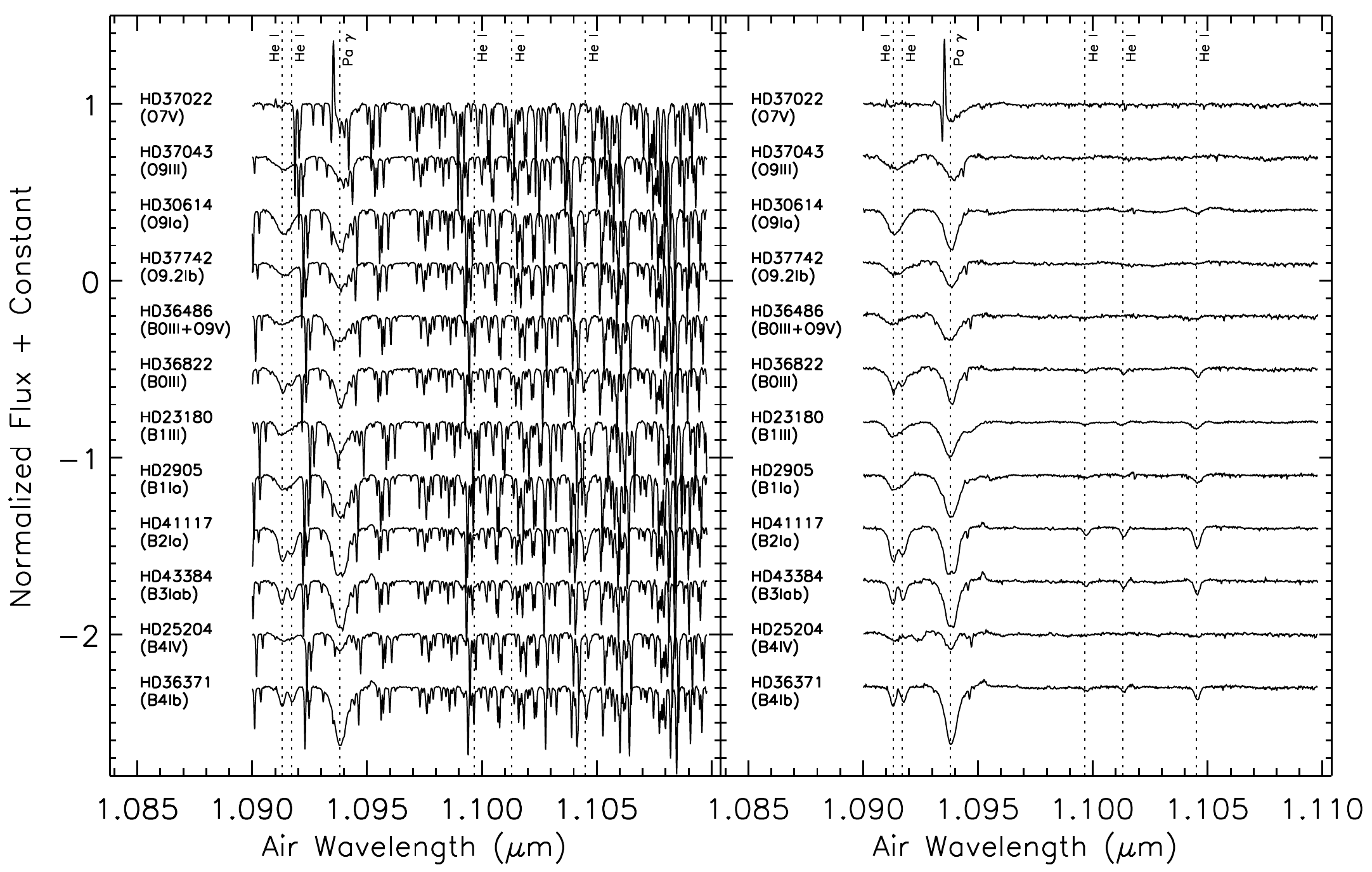}
 \caption{Spectra (order 51) of 12 OB stars prior to ({\it left panel})
 and after telluric correction ({\it right panel}).  Each spectrum is
 shifted so that the line center of Pa\,$\gamma$ lies at 10938.09 \AA.
 Note that the spike feature near Pa\,$\gamma$ in HD 37022 is not caused
 by error in the telluric correction but due to incomplete subtraction
 of the emission line from the surrounding Orion Nebula.}
 \label{fig:OBstars}
\end{figure*}

As another example, the results of telluric correction for several OB
stars are illustrated in Figure \ref{fig:OBstars}.  In the observed
wavelength range 1.09--1.11 \micron, serious absorption by water vapor
makes it difficult to properly trace stellar features such as
Pa\,$\gamma$.  Our method successfully removed the telluric absorption
for all objects, making it possible to trace the entire line profiles of
several lines, including Pa\,$\gamma$.  Moreover, the weak helium
lines---\ion{He}{1} $\lambda\lambda10996.6, 11013.1, 11045.0$---are
barely identifiable in the observed spectra but are clearly identified
for some objects after telluric correction.

%%%%%%%%%%%%%%%%%%%%%%%%%%%%%%%%%%%%%%%%%%
% Discussion
%%%%%%%%%%%%%%%%%%%%%%%%%%%%%%%%%%%%%%%%%%

\section{Discussion}

The aim of this study was to establish a method for correcting telluric
absorption in NIR high-resolution and high-SNR spectra.  The proposed
method uses A0\,V stars or their analogs as telluric standard stars
whose intrinsic stellar lines were carefully removed with the help of
synthetic telluric spectra created using {\tt molecfit} as a reference.
As presented in the previous section, our method works effectively even
for feature-rich G-type stars and can successfully reproduce weak
stellar lines that are barely identifiable in the observed spectrum
prior to removal of heavy telluric absorption.

In the following subsections, we will evaluate the accuracy of our
method by comparing it to other methods and through the use of a newly
developed diagnostic method.  We will also investigate the dependence of
telluric correction on observing conditions, i.e., airmass and time, in
a quantitative manner.

\subsection{Comparison with simple spectral division by early-type stars} \label{sec:comparison_methods}

\begin{figure}[t!]
 \epsscale{1.1}
 \plotone{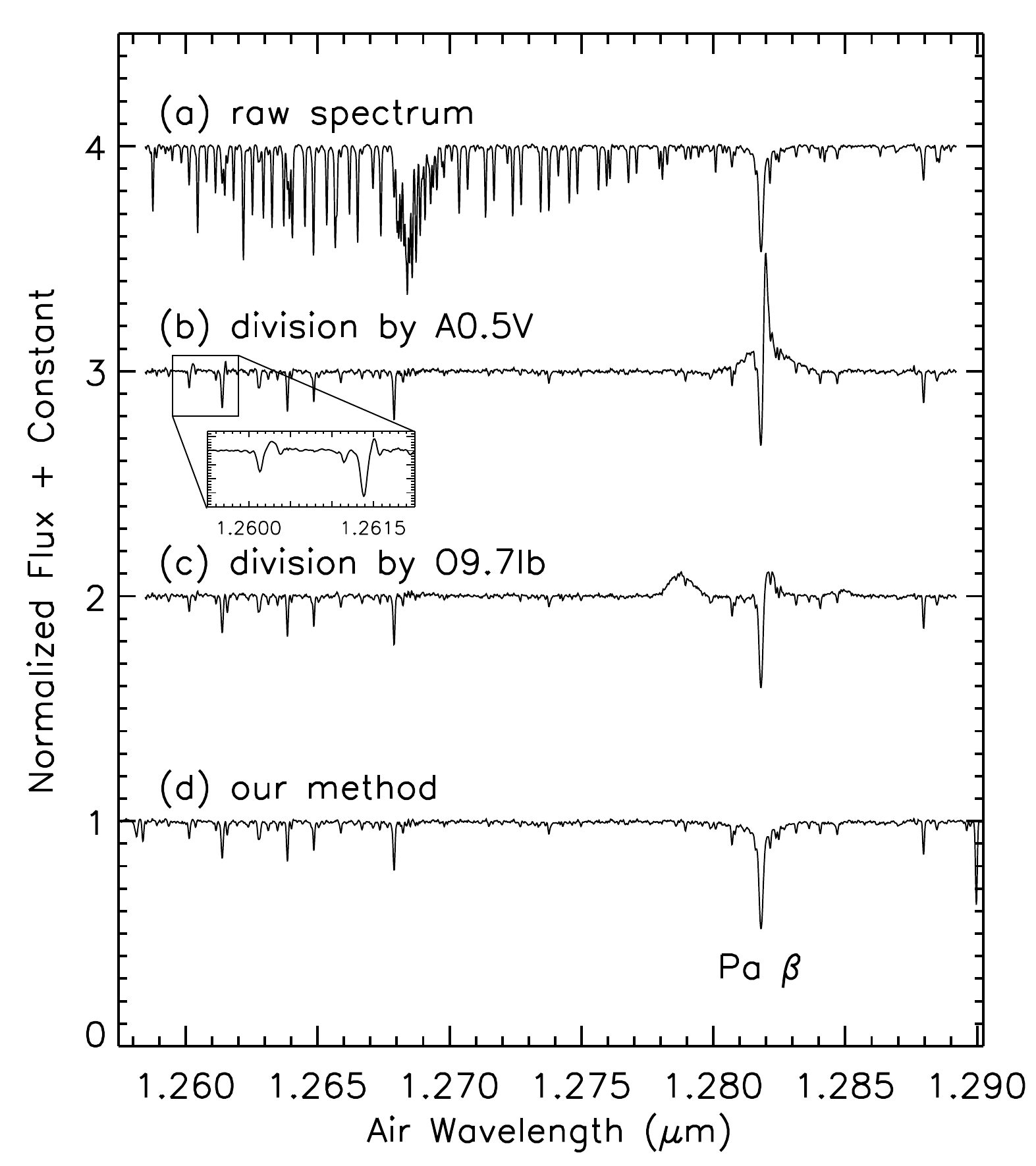}
 \caption{Comparison between observed and corrected spectra of
 $\varepsilon$ Leo (order 44) with three different telluric corrections
 considered: ($b$) division by an A0.5\,V star (21 Lyn), ($c$) division
 by an O9.2\,Ib star (HD 37742), and ($d$) our method.}
 \label{fig:comparison}
\end{figure}

An observed spectrum of a featureless early-type star is sometimes used
to correct telluric absorption without removal of its stellar lines.
Here, we demonstrate some of the limitations of this approach relative
to our method.  Figure \ref{fig:comparison} shows the observed spectrum
of $\varepsilon$ Leo for spectral order 44 divided by that of an A- or
O-type star and a spectrum corrected using our method.  The performance
of the respective approaches are summarized as follows:

1. Spectral division by an A-type star significantly distorts the
Pa\,$\beta$ line.  Furthermore, two emission-like residuals are seen at
$\sim$12,603 \AA\ and $\sim$12,615 \AA\ as a result of \ion{C}{1}
$\lambda\lambda 12601, 12614$ absorption lines of the A-type star,
respectively (see Figure \ref{fig:fitting_intrinsic_lines}).  It is seen
that, in the case of NIR high-resolution and high-SNR spectra, both the
strong hydrogen and weak metal lines of a telluric standard star should
be removed.  From our analysis of WINERED spectra we found that even
A0\,V stars have $\sim$100 metal lines that must be removed at
0.90--1.35 \micron.  The strengths of these metal lines vary
considerably with $v \sin i$, metallicity, chemical composition, etc.,
which prevented us from transforming the spectrum of Vega into that of
another A-type star by changing the scales and widths of the stellar
lines, as required in the method proposed by \cite{2003PASP..115..389V}.

2. O-type stars might serve as better telluric standards than A-type
stars because they have smaller numbers of metal lines.  Moreover, their
high rotational velocities make their absorption lines shallow,
resulting in weak residuals in the target spectrum after telluric
correction.  However, O-type stars have many strong hydrogen and helium
lines that must be fitted and removed.  This is clearly illustrated in
Figure \ref{fig:comparison}($c$), in which there are bumps at
$\sim$1.279 and $\sim$1.282 \micron\ corresponding to \ion{He}{1}
$\lambda\lambda$12784, 12790 and Pa\,$\beta$, respectively, in the
O-type star.  Other problems in using O-type stars as telluric standards
include: (1) their sparse distribution on the celestial sphere compared
to A-type stars, and; (2) the fact that they often have strong emission
lines from surrounding gas.

3. Unlike simple spectral division with A- and O-type stars, our method
does not disturb the spectral components around Pa\,$\beta$.  In
addition, it does not have to account for emission-line-like noises
owing to the weak metal lines found in telluric standard stars.  Our
method is therefore quite effective in correcting the telluric
absorption in NIR high-resolution and high-SNR spectra.  However, our
method does have some disadvantages.  It requires somewhat tedious
procedures involving the manual fitting of many weak metal lines and
some strong hydrogen lines.  One possible way to significantly reduce
this burden is to reapply the fitted intrinsic spectrum
$[I_{\mathrm{std}}^\prime(\lambda) * P(\lambda)]_{\mathrm{fit}}$ to the
observed spectra of the same standard star taken at different times.  By
routinely observing selected A0\,V stars in this manner, it would be
possible to automatically remove their intrinsic stellar lines.  After
several experiments along these lines, we found that a satisfactory
level of removal can be achieved for weak metal lines, but
non-negligible residuals remain around the hydrogen lines, primarily
because of the uncertainty in continuum normalization around such broad
lines; further investigations will be required along these lines.

\subsection{Quality of telluric correction} \label{sec:qual_diag}

We also developed a new diagnostic method for evaluating the quality of
telluric correction.  Here, we denote the fluxes of a
continuum-normalized spectrum before and after telluric correction as
$F(\lambda)$ and $G(\lambda)$, respectively.  If the correction is
perfect, $G(\lambda)$ should be unity at positions where the intrinsic
stellar lines of the target are not present.  The deviation of
$G(\lambda)$ from unity can be used, in addition to the dispersion of
$G(\lambda)$, as a quality indicator of telluric correction.
Furthermore, unless the accuracy of the performed telluric correction is
poor the ratio $F(\lambda)/G(\lambda)$ can be regarded as an
approximation of the atmospheric transmittance $T(\lambda)$.  Therefore,
by comparing $G(\lambda)$ and $F(\lambda)/G(\lambda)$, we can
quantitatively examine the accuracy of telluric correction and its
dependence on atmospheric transmittance.

\begin{figure*}[t!]
 \epsscale{1.1}
 \plotone{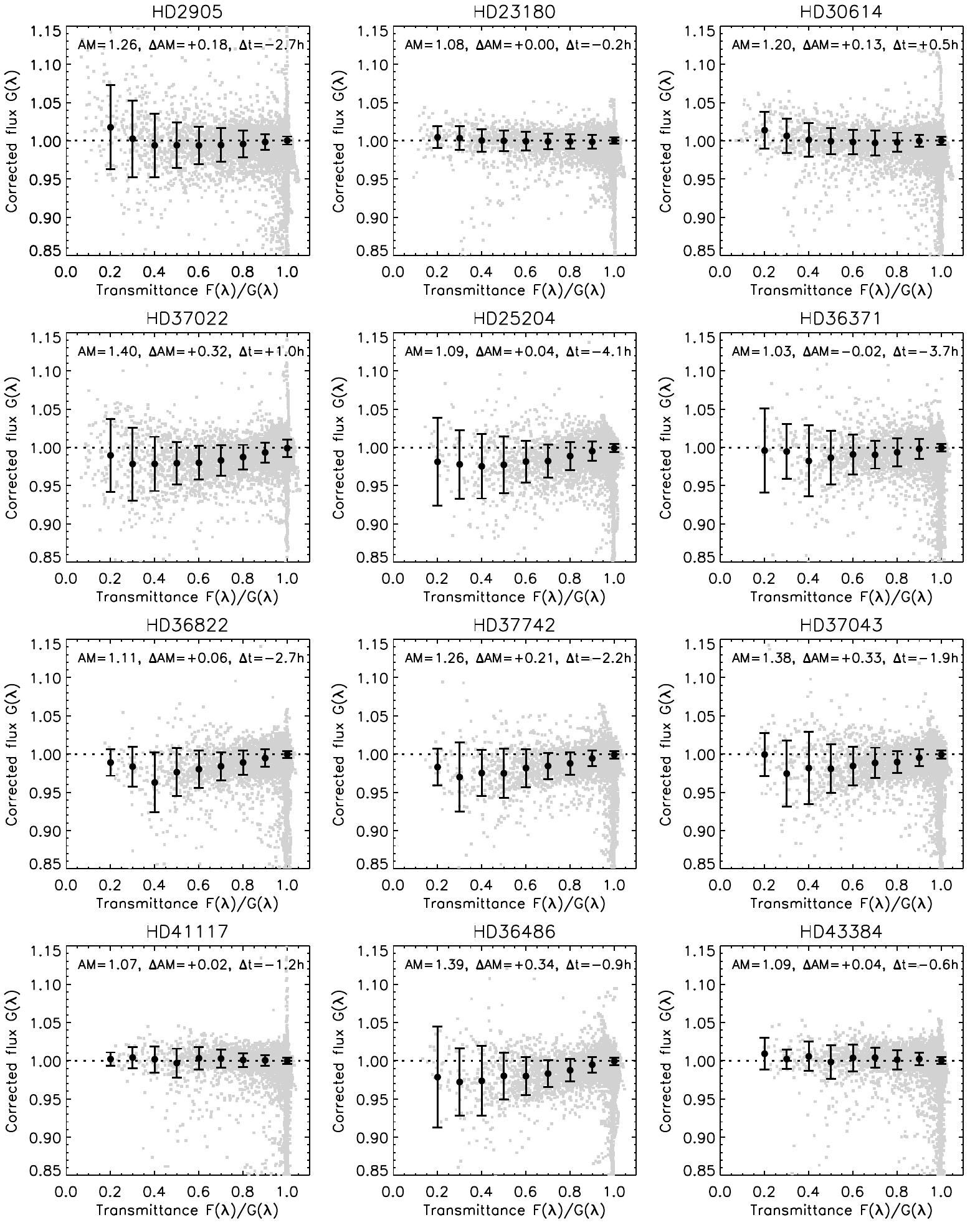}
 \caption{Quality-diagnostic diagrams for 12 OB stars.  Gray dots
 indicate the telluric-corrected flux $G(\lambda)$ plotted against
 $F(\lambda)/G(\lambda)$, a proxy for atmospheric transmittance, for
 each pixel outside of the O$_2$-band region (1.245--1.280 \micron).  The
 circles and error bars indicate, respectively, the sigma-clipped mean
 and standard deviation for each transmittance bin ($\Delta=0.1$).  The
 mean airmass of the target, the difference in airmass from that of the
 telluric standard star, and the time difference are given in each panel.}
 \label{fig:telluric_depth}
\end{figure*}

\begin{figure*}[t!]
 \epsscale{1.1}
 \plotone{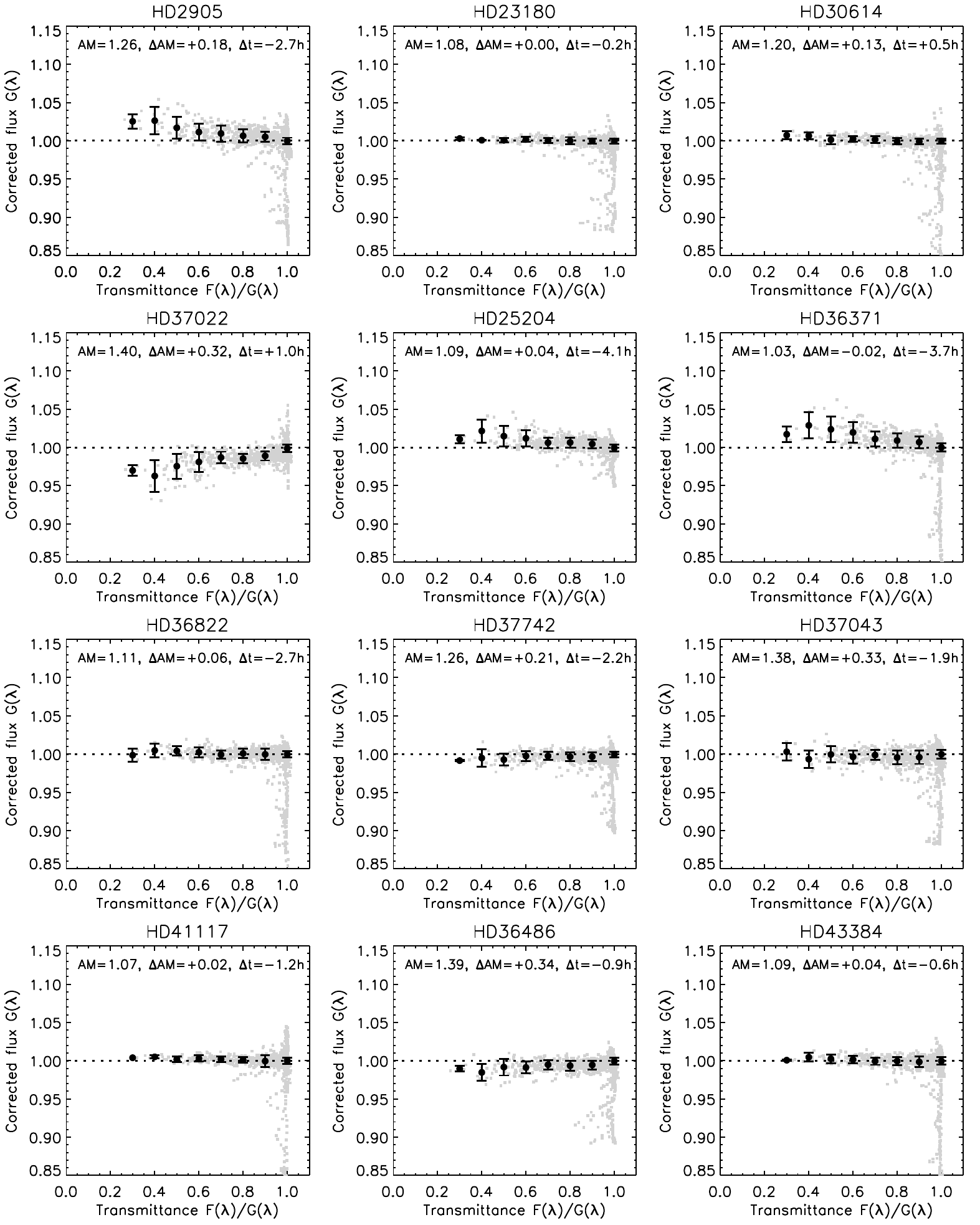}
 \caption{Quality-diagnostic diagrams for 12 OB stars as Figure
 \ref{fig:telluric_depth}, but using data in the O$_2$-band region
 (1.245--1.280 \micron).}  \label{fig:telluric_depth_o2}
\end{figure*}

We applied the above diagnostic method to the 12 OB stars listed in
Table \ref{tab:obslog}.  To check differences among the absorbers, we
separated the star's spectral parts into two groups that are either
affected primarily by water vapor absorption or by molecular oxygen
absorption; the results are illustrated in Figures
\ref{fig:telluric_depth} and \ref{fig:telluric_depth_o2}, respectively.
Note that the wavelength ranges 0.930--0.960, 1.115--1.160, and $>$1.330
\micron\ were not included in our analysis (see \S\ref{sec:results}).
Because intrinsic stellar lines also lead to offsets of $G(\lambda)$
from unity, this diagnostic works better for featureless objects; for
this reason, we excluded $\varepsilon$ Leo from this analysis.  The
scattering seen in Figures \ref{fig:telluric_depth} and
\ref{fig:telluric_depth_o2} is caused by various factors including
random noise owing to SNR, systematic deviations owing to intrinsic
stellar lines, and inaccuracies in continuum normalization.  Note that,
in our case, the SNR is high enough and the especially large vertical
scatter seen at $F(\lambda)/G(\lambda) \sim 1$ mainly reflects the
intrinsic stellar lines of the target.  In addition, stellar features
are expected to give outliers at each level of $F(\lambda)/G(\lambda)$
smaller than one.  To avoid such stellar contamination, we calculated
3-sigma-clipped means and standard deviations for each bin of
$F(\lambda)/G(\lambda)$, which are plotted in the figure.

It is seen from Figure \ref{fig:telluric_depth} that the quality of
telluric correction is particularly good for HD 23180, HD 41117, and HD
43384.  In these cases, the deviation of $G(\lambda)$ from unity is
$\lesssim1$\% and the dispersion of $G(\lambda)$ is $\lesssim2$\%, even
at $F(\lambda)/G(\lambda) \sim0.2$; this indicates that our method works
for atmospheric transmittances as low as $\sim$20\%.  In all these
cases, (1) the airmass differs from that associated with the telluric
standard star by $\lesssim0.05$, and (2) the time differs by $\lesssim
1$ h.  In other words, telluric standard stars should be observed to
fulfill these criteria at the Koyama Astronomical Observatory.
Considering high humidity ($\gtrsim$80\%) and relatively unstable
weather conditions at the Koyama Astronomical Observatory, it may safely
be said that these criteria would be relaxed at sites with better
conditions like Paranal or Mauna Kea.

As seen in Figure \ref{fig:telluric_depth_o2}, the dispersion of
$G(\lambda)$ for molecular oxygen absorption is generally small compared
to that for water vapor absorption.  However, in some cases offsets from
unity are seen in $G(\lambda)$ where the telluric transmittance is low.
We determined that these offsets are primarily caused by the absorption
at 1.268--1.270 \micron, where telluric lines are closely blended with
each other.  This may indicate that changes in line shape as a function
of airmass and time are different for isolated telluric and blended
lines; thus, simple scaling of the standard spectrum optimized for
unblended telluric lines results in offsets of $G(\lambda)$ from unity
for blended lines.  It is remarkable that, even in such extreme regions,
telluric correction can be performed successfully based on observations
fulfilling the above conditions on airmass and time.

\begin{figure}[t!]
 \epsscale{1.1}
 \plotone{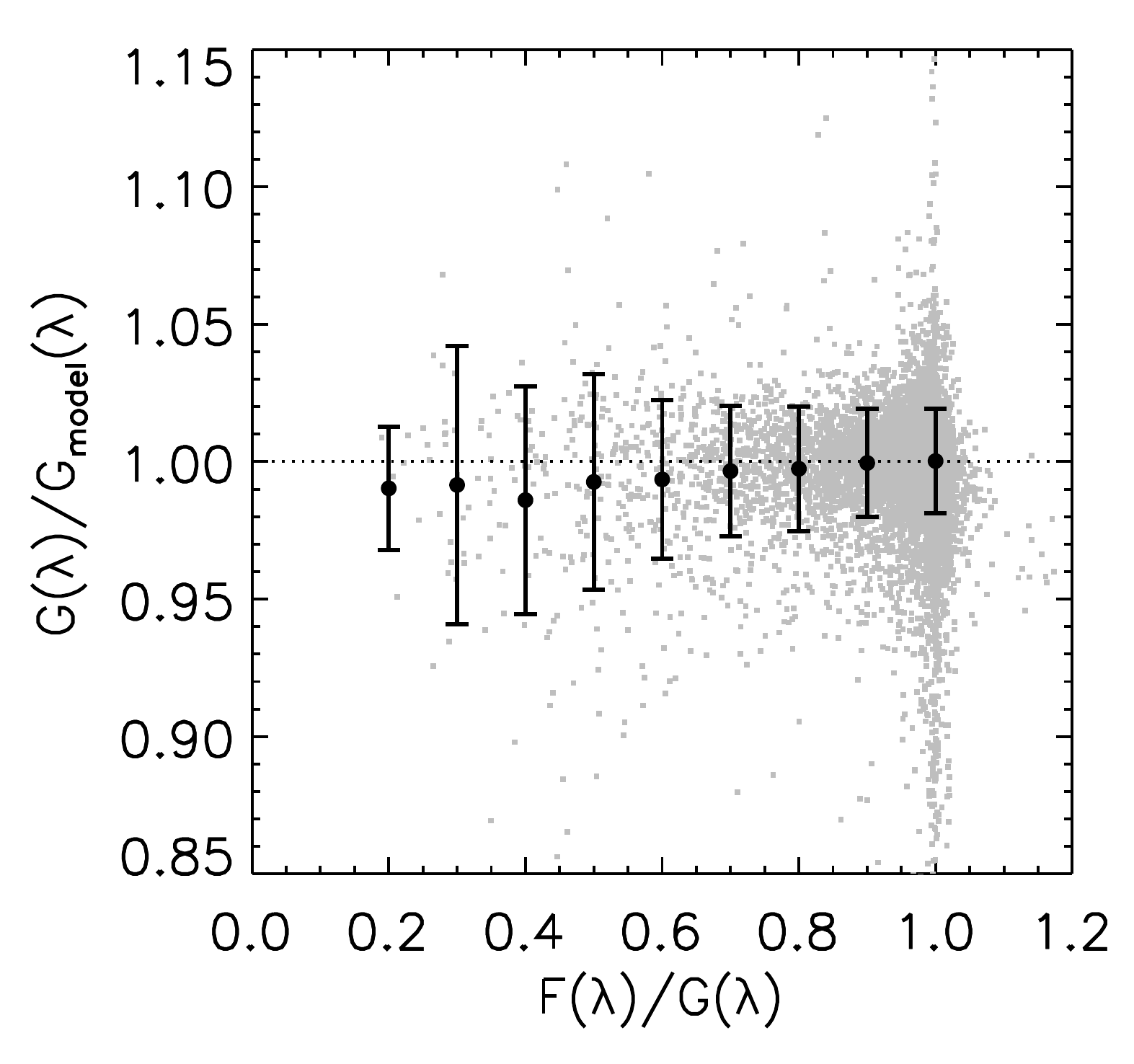}
 \caption{Quality-diagnostic diagram for $\varepsilon$ Leo presented in
 the same manner as in Figure \ref{fig:telluric_depth} but using
 $G(\lambda)/G_\mathrm{model}(\lambda)$ as the vertical axis instead of
 $G(\lambda)$.}  \label{fig:diag_epsleo}
\end{figure}

We also applied our diagnostic method to the G-type giant $\varepsilon$
Leo and obtained the result shown in Figure \ref{fig:diag_epsleo}.
Because $\varepsilon$ Leo has many intrinsic absorption lines, it
produces a non-negligible number of $G(\lambda)$ that deviate from unity
even under perfect telluric correction.  Therefore, we plotted
$G(\lambda)/G_\mathrm{model}(\lambda)$ on the vertical axis instead of
$G(\lambda)$, where $G_\mathrm{model}(\lambda)$ is a model spectrum of
$\varepsilon$ Leo created by {\tt SPTOOL} (see \S\ref{sec:results}).  As
seen in the figure, the difference between the corrected and model
spectra as measured by the standard deviation of
$G(\lambda)/G_\mathrm{model}(\lambda)$ is less than 5\% at $0.2 \lesssim
F(\lambda)/G(\lambda) \lesssim 1$.  The scatter is larger than the best
cases in Figure \ref{fig:telluric_depth} but similar to those cases with
relatively large scatters.  Note that, as the scatter in Figure
\ref{fig:diag_epsleo} includes errors in the model spectrum, the scatter
is an upper limit of the accuracy of the telluric correction applied to
$\varepsilon$ Leo.  In fact, the standard deviation in Figure
\ref{fig:diag_epsleo} is larger at $F(\lambda)/G(\lambda)=1$ than those
in Figures \ref{fig:telluric_depth} and \ref{fig:telluric_depth_o2} and
increases slowly with decreasing $F(\lambda)/G(\lambda)$, which
indicates the significant contribution of error sources other than the
telluric correction.  This confirms that our method can produce telluric
corrections with an accuracy of 5\% or better for G-type stars.

\subsection{Comparison with {\tt molecfit}}

Comparing our method with {\tt molecfit} should be useful for readers to
select their strategies for observation because very accurate telluric
correction may not be needed for many scientific cases, for which
observing telluric standard stars may be expensive in terms of telescope
time.  For medium-resolution spectra, \cite{2015A&A...576A..78K} compare
a classical method based on telluric standard stars and the telluric
correction by {\tt molecfit} using VLT/X-Shooter spectra.  Here, we
compare the two methods for NIR high-resolution spectra by using WINERED
spectra ($R \sim 28,000$).

We used the spectra of HD 23180 (B1\,III; see Table \ref{tab:obslog})
for the test. To optimize the {\tt molecfit} parameters, we first
obtained best-fit parameters from the fitting of the telluric standard
star (7 Cam) following the procedures written in \S\ref{sec:molecfit}
and put them as initial values in the fitting of HD 23180.  Because the
intrinsic stellar lines of HD 23180 are strong and wide, they could be
distinguished from telluric absorption lines and masked to improve the
fitting.

\begin{figure}[t!]
 \epsscale{1.1}
 \plotone{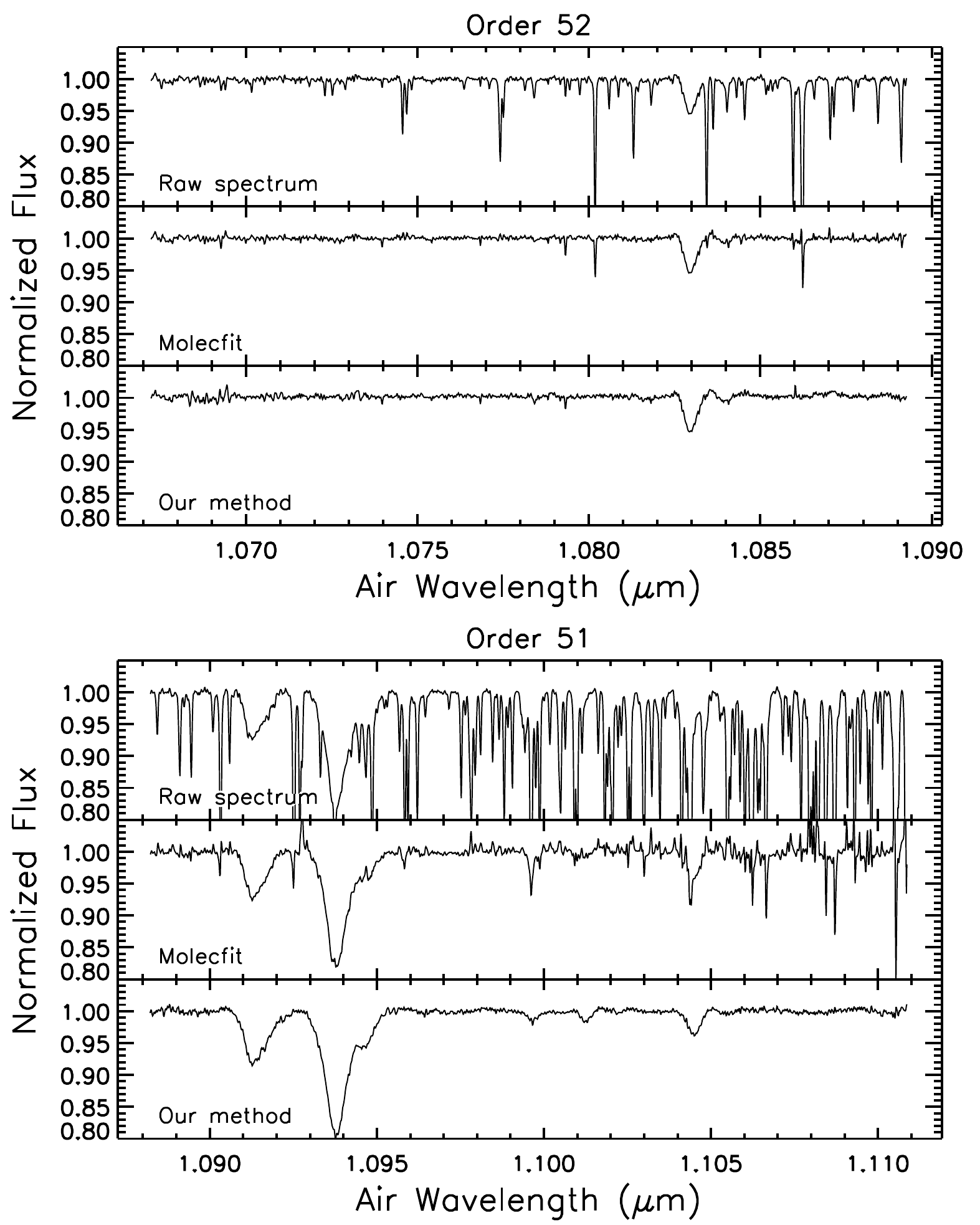}
 \caption{Telluric correction of HD 23180 by our method and {\tt
 molecfit} for the orders 52 ({\it upper panel}) and 51 ({\it lower
 panel}).  In each panel, the raw spectrum, which is not corrected for
 telluric absorption, is shown for reference.}
 \label{fig:compare_molecfit}
\end{figure}

In Figure \ref{fig:compare_molecfit}, the resultant telluric correction
of HD 23180 by {\tt molecfit} is compared to that with our method for
the orders 52 and 51.  {\tt Molecfit} can nicely correct telluric
absorption overall, but the fitting residual becomes non-negligible at
the wavelength region where absorption lines are strong and blended.
Although we tried fine-tuning parameters as much as possible, those
fitting residuals could not be removed.  This must be partly due to
imperfect modeling of the instrumental profile and wavelength solution
in our molecfit running, but there may be another possibility.  Those
strong absorption lines may be saturated in reality but residual flux
shows up in their core owing to instrumental smearing.  If that is the
case, Figure \ref{fig:compare_molecfit} may imply that our method can
correct even for moderately saturated telluric lines with an accuracy of
the order of 2\% or better of the continuum , while
\cite{2015A&A...576A..77S} report that {\tt molecfit} can reach such an
accuracy in correcting unsaturated lines.

\begin{figure}[t!]
 \epsscale{1.1}
 \plotone{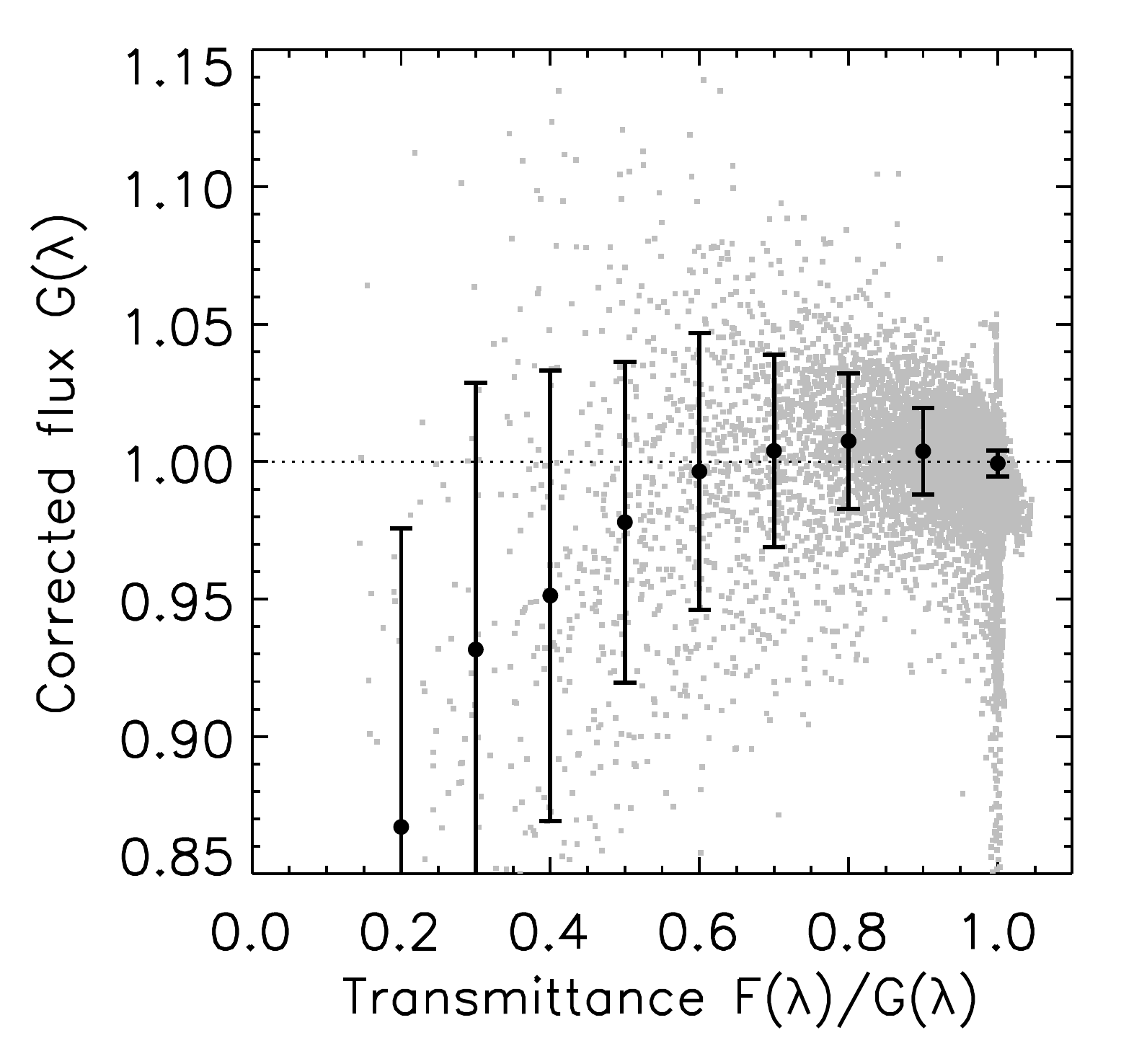}
 \caption{Quality-diagnostic diagram of HD 23180 whose telluric
 absorption is corrected by {\tt molecfit}.  Pixels outside of the
 O$_2$-band region (1.245--1.280 \micron) are used in the plot as in
 Figure \ref{fig:telluric_depth}.}  \label{fig:diag_HD23180}
\end{figure}

The diagnostic method introduced in \S\ref{sec:qual_diag} can be applied
to spectra corrected by {\tt molecfit}, which enables a direct
comparison between the two methods in a quantitative manner.  Figure
\ref{fig:diag_HD23180} shows the result for the {\tt molecfit}-corrected
spectrum of HD 23180.  Telluric correction with {\tt molecfit} results
in larger scattering than our method (see Figure
\ref{fig:telluric_depth}) and the deviation of the corrected flux
$G(\lambda)$ from unity increases as the transmittance decreases;
$G(\lambda)$ becomes noticeably low ($<0.95$) at the transmittance
$F(\lambda)/G(\lambda) < 0.5$.  This quantitative analysis is consistent
with the qualitative analysis above (see Figure
\ref{fig:compare_molecfit}).  Similar trends are also found for all OB
stars, for which we tried the same test.  As discussed above, the
deviation of $G(\lambda)$ from unity at low transmittance may be due to
saturated absorption lines, while {\tt molecfit} works fine for
unsaturated absorption lines.

\begin{figure}[t!]
 \epsscale{1.1}
 \plotone{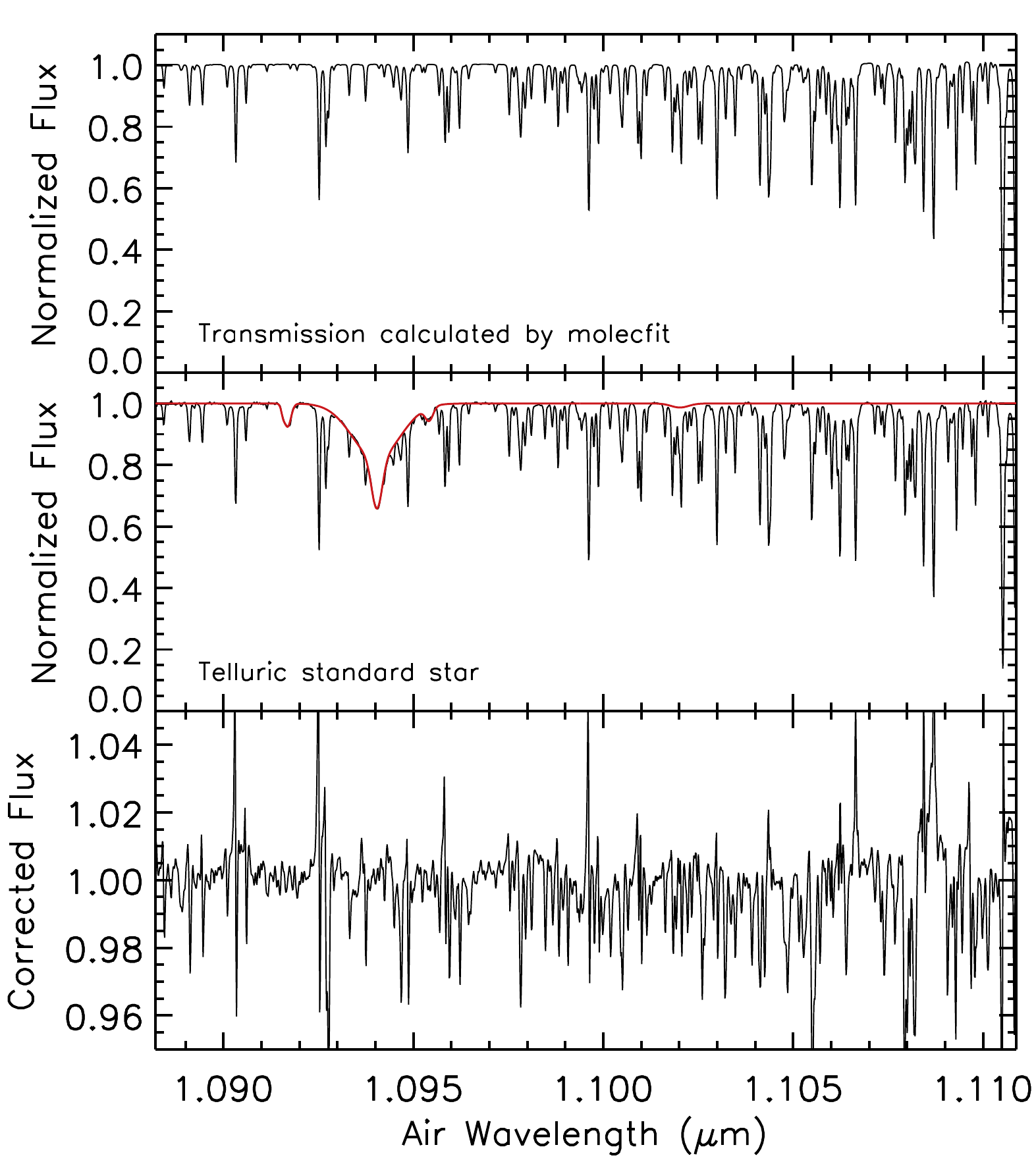}
 \caption{{\it Upper panel}: best-fit transmission curve created by {\tt
 molecfit} from HD 23180 (order 51).  {\it Middle panel}: spectrum of
 the telluric standard star 7 Cam (black) and the removed intrinsic
 stellar lines (red).  {\it Lower panel}: after the telluric correction
 of the best-fit transmission curve ({\it upper panel}) using the
 telluric standard star spectrum ({\it middle panel}) by our method.}
 \label{fig:ref_test}
\end{figure}

Another interesting test is to use the best-fit transmission curve
created by {\tt molecfit} from a science target spectrum as a
pseudo-target spectrum and apply our method to it.  In the ideal case,
this will produce a straight line at unity.  Deviations from unity will
arise from any one of the following: incompleteness in the {\tt
molecfit} modeling, inaccuracies in our method, and noise.  The spectrum
of HD 23180 (order 51) was used for this test and the result is shown in
Figure \ref{fig:ref_test}.  As can be seen from the lower panel in the
figure, the deviations of $\gtrsim 1$\% are mainly seen where telluric
absorption is strong, indicating differences between the two methods at
those regions.  Significant deviations are not confirmed at the
wavelength region where intrinsic lines of a telluric standard star were
removed.  In addition to such high-frequency deviations, low-frequency
deviations with an amplitude of $<1$\% are also confirmed in the
spectrum.  This is due to the inaccuracies in the continuum
normalization of a telluric standard star's spectrum, which is almost
inevitable at the wavelength region where telluric absorption lines are
crowded.  As \cite{2015A&A...576A..78K} pointed out, inaccuracies in
continuum normalization is one of the disadvantages of classical methods
using telluric standard star.  Although it is beyond the scope of this
paper, extracting low-frequency function from this spectrum and use it
to improve the continuum normalization of the telluric standard star
could be a good idea to overcome the normalization problem.

To summarize, {\tt molecfit} effectively corrects unsaturated telluric
lines with the accuracy which many science cases require.  On the other
hand, the above tests indicate that our method is applicable even to
strong telluric lines with moderate saturation.  If the target line is
very faint ($\lesssim$ a few \% in depth) and blended with strong
telluric lines (see, e.g., \ion{C}{1} lines at $\sim$1.26 \micron\ in
Figure \ref{fig:fitting_intrinsic_lines}; \ion{He}{1} lines at
$\sim$1.10 \micron\ in Figure \ref{fig:OBstars}), our method could be an
efficient way to correct telluric absorption.

\subsection{Effects of separations in airmass and time}

\begin{deluxetable*}{lccccc}
\tablecaption{Observing Conditions \label{tab:conditions}}
\tablehead{
\colhead{Name} & \colhead{Standard} & \multicolumn{2}{c}{Difference from
 Standard} & \multicolumn{2}{c}{$\sigma_{50}$} \\
 \cmidrule(r){3-4}\cmidrule{5-6}
 \colhead{} & \colhead{} & \colhead{Airmass} & \colhead{Time (hour)} & \colhead{H$_2$O} & \colhead{O$_2$}
 }
 \startdata
 HD 2905           & 7 Cam  & $+0.18$ & $-2.7$ & 0.0298 & 0.0107 \\
 HD 23180          & 7 Cam  & $+0.00$ & $-0.2$ & 0.0131 & 0.0033 \\
 HD 30614          & 7 Cam  & $+0.13$ & $+0.5$ & 0.0170 & 0.0039 \\
 HD 37022          & 7 Cam  & $+0.32$ & $+1.0$ & 0.0278 & 0.0132 \\
 HD 25204          & 21 Lyn & $+0.04$ & $-4.1$ & 0.0368 & 0.0105 \\
 HD 36371          & 21 Lyn & $-0.02$ & $-3.7$ & 0.0353 & 0.0132 \\
 HD 36822          & 21 Lyn & $+0.06$ & $-2.7$ & 0.0313 & 0.0067 \\
 HD 37742          & 21 Lyn & $+0.21$ & $-2.2$ & 0.0319 & 0.0065 \\
 HD 37043          & 21 Lyn & $+0.33$ & $-1.9$ & 0.0318 & 0.0085 \\
 HD 41117          & 21 Lyn & $+0.02$ & $-1.2$ & 0.0188 & 0.0040 \\
 HD 36486          & 21 Lyn & $+0.34$ & $-0.9$ & 0.0304 & 0.0076 \\
 HD 43384          & 21 Lyn & $+0.04$ & $-0.6$ & 0.0221 & 0.0048 \\
 $\varepsilon$ Leo & 21 Lyn & $-0.03$ & $+1.7$ & 0.0391\tablenotemark{a} & 0.0091\tablenotemark{a}\\
 \enddata
 \tablenotetext{a}{$\sigma_{50}$ values for $\varepsilon$ Leo were
 estimated in a different manner than for other objects (OB stars)
 through the use of a model spectrum created for this kind of G-type
 star (see text).}
\end{deluxetable*}

\begin{figure*}[t!]
 \epsscale{1.1}
 \plotone{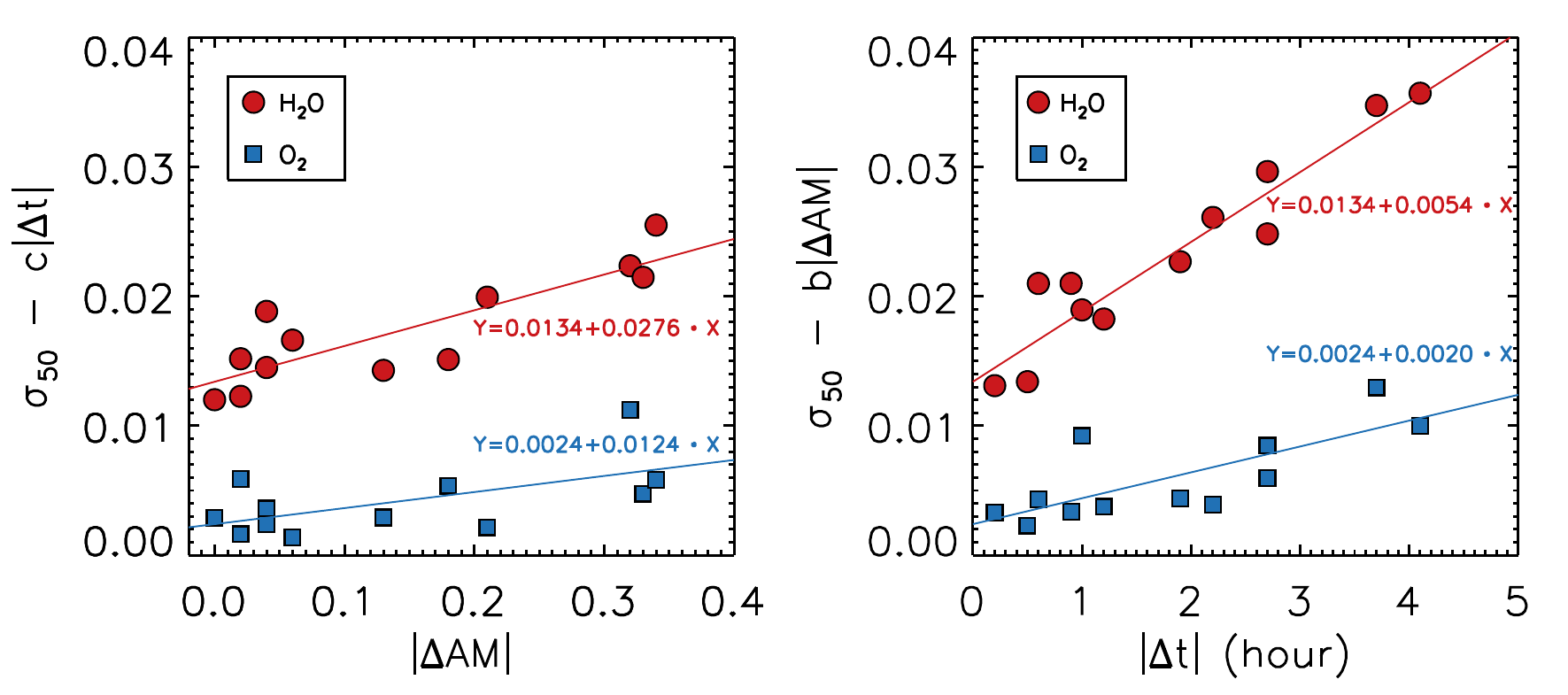}
 \caption{$\sigma_{50}$; standard deviations of the corrected flux at an
 atmospheric transmittance of 50\% (see text for further details)
 plotted against the difference in airmass ({\it left}) and time ({\it
 right}).  In both panels, the contribution of other parameters to
 $\sigma_{50}$ are removed using Eq.\,(\ref{eq:linfit}).  The red
 circles and blue boxes indicate, respectively, the results for
 absorption by water vapor and molecular oxygen.  The results of
 multiple linear regression are indicated with solid lines.}
 \label{fig:sigma}
\end{figure*}

Here, we discuss in more detail how observing conditions, i.e.,
differences in airmass and time between a target and a standard star,
affect the accuracy of telluric correction.  We use $\sigma_{50}$ to
denote the standard deviation of $G(\lambda)$ at $0.45 <
F(\lambda)/G(\lambda) < 0.55$, which indicates the accuracy of the
correction for pixels for which telluric absorption is as deep as 50\%.
The $\sigma_{50}$ values measured for the 12 OB stars and $\varepsilon$
Leo are listed in Table \ref{tab:conditions}, along with the differences
in airmass and time from those associated with the standard stars.
Using the data for the 12 OB stars, we performed multiple linear
regression assuming the following model:
\begin{equation}
 \sigma_{50} = a + b\,|\Delta \mathrm{AM}| + c\,|\Delta t|, \label{eq:linfit}
\end{equation}
where $|\Delta \mathrm{AM}|$ and $|\Delta t|$ are the absolute values of
the difference in airmass and time in hours, respectively.
Least-squares fitting produces the coefficients $(a,b,c)=(0.0134,
0.0276, 0.0054)$ for water vapor absorption and $(0.0024, 0.0124,
0.0020)$ for molecular oxygen absorption.  This indicates that, even in
the case of water vapor absorption, $\sigma_{50} \lesssim 2\%$ can be
maintained if $|\Delta \mathrm{AM}| \lesssim 0.05$ and $|\Delta t|
\lesssim 1$ h, which is consistent with the discussion in the previous
subsection.  In Figure \ref{fig:sigma}, $\sigma_{50} - c\,|\Delta t|$
and $\sigma_{50} - b\,|\Delta \mathrm{AM}|$ are plotted against $|\Delta
\mathrm{AM}|$ and $|\Delta t|$, respectively, to clarify the dependence
of $\sigma_{50}$ on each parameter separately.
 
We also performed partial correlation analysis to remove interdependence
among the parameters.  The partial correlation coefficients between
$\sigma_{50}$ and $|\Delta \mathrm{AM}|$ were calculated using the IDL
procedure {\tt p\_correlate.pro} as follows:
\begin{eqnarray}
 r(\sigma_{50,\mathrm{H_2O}}, |\Delta \mathrm{AM}|; |\Delta t|) &=& 0.849\ (< 4.8 \times 10^{-4}), \\
 r(\sigma_{50,\mathrm{O_2}}, |\Delta \mathrm{AM}|; |\Delta t|) &=& 0.591 \ (< 0.043),
\end{eqnarray}
where $r(X,Y;Z)$ denotes the partial correlation coefficient between $X$
and $Y$ holding $Z$ fixed, $\sigma_{50,\mathrm{H_2O}}$ and
$\sigma_{50,\mathrm{O_2}}$ are the $\sigma_{50}$ for water vapor and
molecular oxygen absorption, respectively, and the parenthesis in the
right-hand side indicates the $p$-value.  Thus, at a significance level
of 0.05 both $\sigma_{50,\mathrm{H_2O}}$ and $\sigma_{50,\mathrm{O_2}}$
positively correlate with $|\Delta \mathrm{AM}|$.  This result leads to
the naturally expected conclusion that the accuracy of telluric
correction improves when the airmass difference between the target and
telluric standard is reduced.  Note, however, that this holds even if a
correction for airmass difference based on Beer's law is performed, as
is done in our method.  Next, the partial correlation coefficients
between $\sigma_{50}$ and $|\Delta t|$ were calculated as
\begin{eqnarray}
 r(\sigma_{50,\mathrm{H_2O}}, |\Delta t|; |\Delta \mathrm{AM}|) &=& 
 0.951 \ (< 2.0 \times 10^{-6}), \\
 r(\sigma_{50,\mathrm{O_2}}, |\Delta t|; |\Delta \mathrm{AM}|) &=& 
 0.751 \ (< 4.9 \times 10^{-3}).
\end{eqnarray}
Although positive correlation is confirmed at a significance level of
$5\times10^{-3}$ in both cases, the correlation coefficient is obviously
larger for water vapor than for molecular oxygen.  This is probably
because the column density of water vapor along a given line of sight is
much more sensitive to temperature and variable over time than the
column density of molecular oxygen.  Besides, the azimuthal dependence
of water vapor concentration may be attributed to the larger correlation
coefficient while molecular oxygen is expected to be almost constant.
Our result suggests that, at least at the Koyama Astronomical
Observatory, where the weather condition can change fairly rapidly, the
time and azimuthal variation of the telluric absorption lines cannot be
fully adjusted by simply changing the variable $\beta$ in
Eq.\,(\ref{eq:beer}).  Thus, in NIR spectroscopic observation it is of
great importance to observe a telluric standard star before or after
observing the target object in either case, preferably within one hour.

%%%%%%%%%%%%%%%%%%%%%
% Summary
%%%%%%%%%%%%%%%%%%%%%

\section{Summary}

We developed a telluric correction method for NIR high-resolution
spectra observed using the WINERED spectrograph.  The proposed method
uses an A0\,V star or its analog as a telluric standard star.  Because
careful removal of intrinsic stellar lines of the telluric standard star
is crucial in the case of high-resolution and high-quality spectra, we
developed a manual process for doing so involving the use of a synthetic
telluric spectrum created using {\tt molecfit}.  By removing the many
weak metal lines identified in the spectrum of an A0\,V star, our method
can successfully correct telluric absorption without disturbing the
intrinsic spectral features of targets including feature-rich G-type
stars. 

We further developed a new diagnostic method to evaluate the accuracy of
telluric correction.  From the application of this diagnostic to 12 OB
stars whose spectra were obtained using WINERED at the Koyama
Astronomical Observatory, we found that we could obtain telluric
correction for spectral parts for which the atmospheric transmittance is
as low as 20\% with accuracies better than 2\% if the following
conditions are fulfilled: (1) the difference in airmass between the
target and the telluric standard star is $\lesssim 0.05$, and; (2) that
in time is $\lesssim 1$ h.  Although readers should take care that the
result is based on a low number statistics and heavily dependent on the
observing site, it would be used as a guide at other sites.  Given that
the humidity tends to be high at the Koyama Astronomical Observatory
($\sim$80\% on the two nights discussed in this paper), the above
conditions may be relaxed for other observation sites with better
weather conditions.  We also applied our diagnostic method to the G-type
giant $\varepsilon$ Leo and found that the reproduced spectrum matched a
model spectrum with better than 5\% accuracy.  Comparison with the
telluric correction by {\tt molecfit} implies that our method may be
applicable to strong telluric absorption lines which are moderately
saturated.  The accuracy of telluric correction depends on both the
difference in airmass and time, with the latter appearing to have a
particularly large impact for water vapor absorption, probably because
the time variability of water vapor and the corresponding spectral
change cannot be fully reproduced by simply scaling a reference
spectrum.  Consequently, minimizing the difference in time between the
target and the telluric standard star is of particular importance in NIR
spectroscopic observation.

%%%%%%%%%%%%%%%%%%%%%
% acknowledgments
%%%%%%%%%%%%%%%%%%%%%

\acknowledgments

We are grateful to the staff of the Koyama Astronomical Observatory for
their support during our observation.  We would like to acknowledge the
helpful comments made by the anonymous referee.  This study is
financially supported by JSPS KAKENHI (Grant Numbers 16684001, 20340042,
21840052, and 26287028), and MEXT Supported Program for the Strategic
Research Foundation at Private Universities, 2008–2012 (No. S0801061)
and 2014–2018 (No. S1411028).  K.F. is supported by JSPS Grant-in-Aid
for Research Activity Start-up (Grant Number 16H07323).
S.H. acknowledges the support from JSPS through Grant-in-Aid for JSPS
Fellows (Grant Number 13J10504).  N.K. is supported by JSPS-DST under
the Japan-India Science Cooperative Programs during 2013--2015 and
2016--2018.

%%%%%%%%%%%%%%%%%%%%%
% appendix
%%%%%%%%%%%%%%%%%%%%%

\appendix

As described in \S\ref{sec:methodology}, the observed
continuum-normalized spectrum $O^\prime(\lambda)$ can be written as
\begin{equation}
 O^\prime(\lambda) = [I^\prime(\lambda) \cdot T(\lambda)] * P(\lambda),
\end{equation}
where $I(\lambda)$ is the intrinsic spectrum of the target star,
$T(\lambda)$ is the telluric absorption spectrum, $P(\lambda)$ is the
instrumental profile, and the asterisk and prime symbol denote
convolution and continuum normalization, respectively.  In spectral
analysis, this equation is often regarded as
\begin{equation}
 O^\prime(\lambda) \sim [I^\prime(\lambda) * P(\lambda)] \cdot [T(\lambda) * P(\lambda)].
\end{equation}
Then, dividing $O^\prime(\lambda)$ by the factor $[T(\lambda) *
P(\lambda)]$ derived from either the spectrum of the telluric standard
star or a synthetic telluric spectrum, we can extract the
(instrumentally smeared) target spectrum $[I^\prime(\lambda) *
P(\lambda)]$.  This procedure is usually considered to be the ``telluric
correction.''  However, the above two equations differ mathematically,
and the difference between $[I^\prime(\lambda) * P(\lambda)]$ and
$O^\prime(\lambda)/[T(\lambda) * P(\lambda)]$ cannot be neglected in
some cases.  Here, we discuss this problem in a quantitative manner
using simple simulations.

The settings for our simulation are summarized as follows.  We created a
synthetic telluric spectrum $T(\lambda)$, using the radiative transfer
code LBLRTM (\citealt{2005JQSRT..91..233C}).  The spectral resolution of
$T(\lambda)$ was set to be very high ($R \sim 500,000$) to enable
simulation of intrinsic telluric absorption with avoiding instrumental
smearing effects.  A Gaussian curve of width corresponding to the
spectral resolution of WINERED ($R \sim 28,000$) was adopted as the
instrumental profile $P(\lambda)$, and convolution was performed using
the IDL script {\tt gaussfold.pro}.  Starting out with artificial
absorption lines with Gaussian profiles that were isolated in the
stellar spectrum $I^\prime(\lambda)$ but blended with telluric
absorption lines, we then simulated several cases by varying the width
of the Gaussian profile and the degree of blending.

\begin{figure*}[t!]
 \epsscale{1.0}
 \plotone{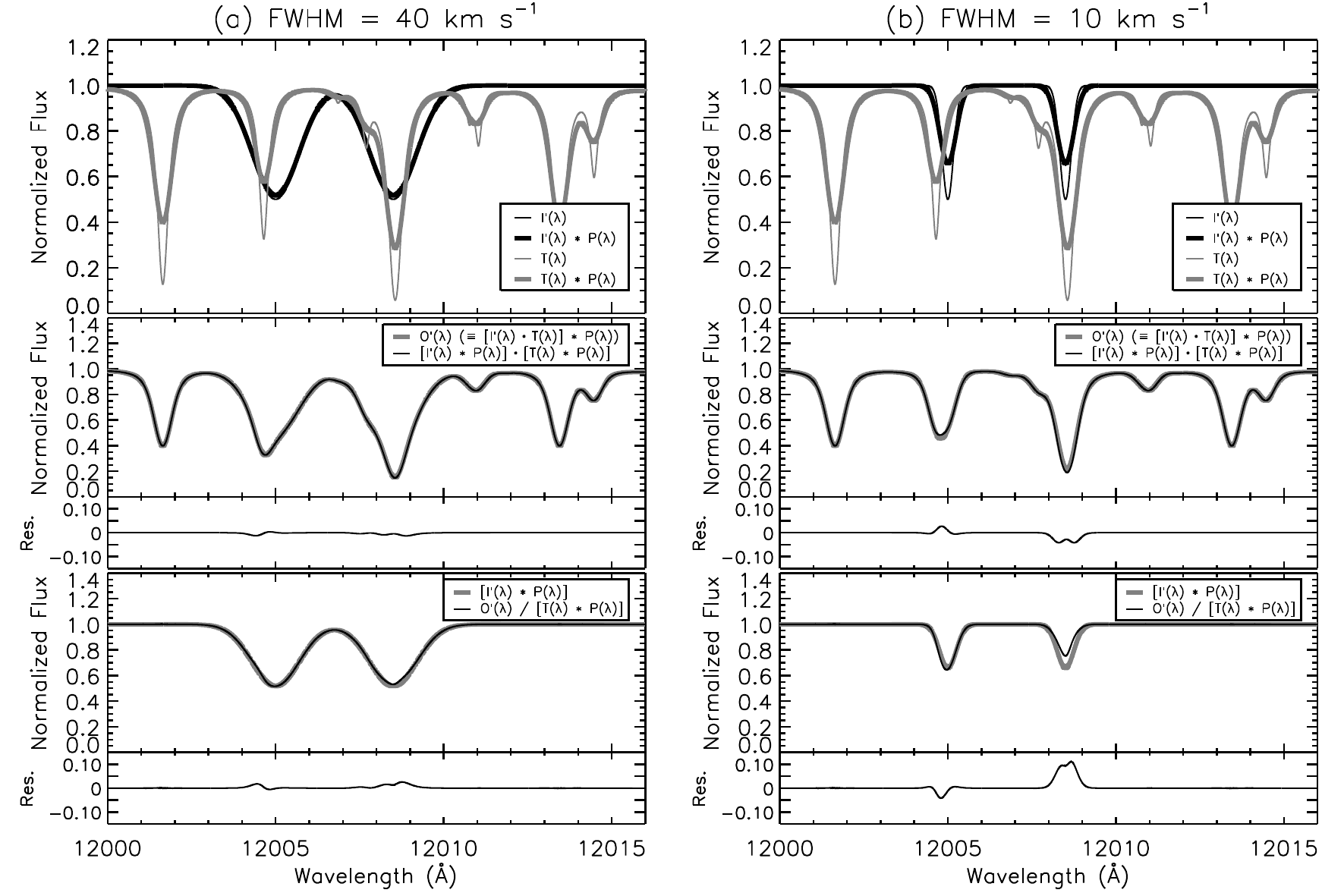}
 \caption{Spectral simulations on the effect of instrumental smearing
 under telluric absorption.  {\it Left:} Simulation for a stellar line
 width of 40 km~s$^{-1}$.  In the top panel, the stellar and telluric
 spectra are indicated by thin black and gray lines, respectively.
 Their instrumentally smeared spectra are indicated by thick lines.  In
 the middle panel, the observed spectrum $O^\prime(\lambda)$ and the
 approximated spectrum $[I^\prime(\lambda) * P(\lambda)] \cdot
 [T(\lambda) * P(\lambda)]$ are indicated by thick gray and black lines,
 respectively.  The residual of the two spectra is also shown in the
 lower part of the middle panel.  In the bottom panel, the true stellar
 spectrum $[I^\prime(\lambda) * P(\lambda)]$ and the reproduced spectrum
 $O^\prime(\lambda)/[T(\lambda) * P(\lambda)]$ are indicated with thick
 gray and thin black lines, respectively.  The residual of the two
 spectra is also shown in the lower part.  {\it Right:} Spectral
 simulations, as in the left panels, but for a stellar line width of 10
 km~s$^{-1}$.}  \label{fig:gaussian_test}
\end{figure*}

\begin{figure*}[t!]
 \epsscale{1.0}
 \plotone{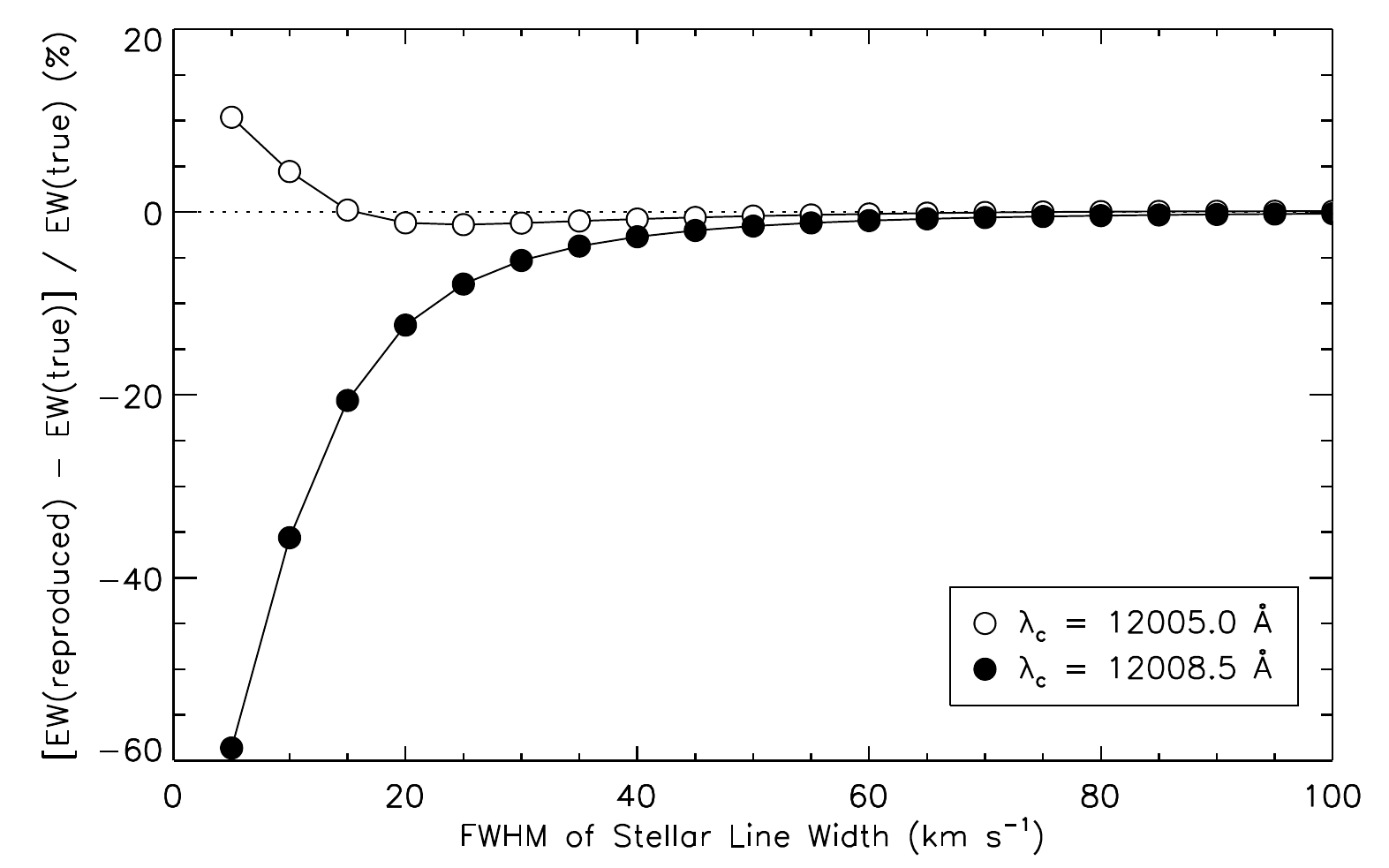}
 \caption{Calculated relative difference in equivalent width between the
 true spectrum ($[I^\prime(\lambda) * P(\lambda)]$) and the reproduced
 spectrum obtained by telluric correction
 ($O^\prime(\lambda)/[T(\lambda) * P(\lambda)]$) as a function of
 stellar line width.  Open and filled circles indicate, respectively,
 cases in which stellar lines are half or fully blended with telluric
 absorption lines.  The results are not significantly affected by the
 stellar line depth.}  \label{fig:ew_dependence}
\end{figure*}

An example of this simulation is shown in Figure
\ref{fig:gaussian_test}, in which two artificial stellar lines are
located at around 12005.0 and 12008.5 \AA, respectively; the former is
blended with the telluric line at 12004.7 \AA\ but separated by about
the width of the instrumental profile, while the latter is completely
blended with the strong telluric line at the same wavelength, 12008.5
\AA.  As is seen in the figure, when the full width at half maximum
(FWHM) of artificial stellar lines is 40 km~s$^{-1}$, they are almost
fully resolved and minimally affected by instrumental smearing.
Correspondingly, the difference between $[I^\prime(\lambda) \cdot
T(\lambda)] * P(\lambda)$ and $[I^\prime(\lambda) * P(\lambda)] \cdot
[T(\lambda) * P(\lambda)]$ is negligible unless the SNR of the spectrum
is very high.  By contrast, when the FWHM of the artificial stellar
lines is 10 km~s$^{-1}$, they are not resolved and are instead smeared
out by convolution with $P(\lambda)$.  As a result, the difference
between $[I^\prime(\lambda) \cdot T(\lambda)] * P(\lambda)$ and
$[I^\prime(\lambda) * P(\lambda)] \cdot [T(\lambda) * P(\lambda)]$ is
non-negligible and increases as the blending between the stellar and
telluric absorption lines increases.  Figure \ref{fig:ew_dependence}
shows the calculated relative difference in equivalent width between the
true stellar profile $[I^\prime(\lambda) * P(\lambda)]$ and the
reproduced profile $O^\prime(\lambda) / [T(\lambda) * P(\lambda)]$ as a
function of stellar line width.  The difference becomes significant if
the line width is comparable to or narrower than the instrumental
resolution and the blending between the stellar and telluric absorption
lines increases.  These results depend on the depth of the contaminating
telluric lines but are not significantly affected by the depth of
stellar lines in $I^\prime(\lambda)$.

To summarize, the spectrum produced by the telluric correction
($O^\prime(\lambda) / [T(\lambda) * P(\lambda)]$) will diverge
significantly from the target's true spectrum ($[I^\prime(\lambda) *
P(\lambda)]$) when the following two conditions are fulfilled: (1) the
intrinsic width of stellar lines is comparable to or narrower than the
instrumental resolution, and (2) the stellar lines are significantly
blended with telluric absorption.  In such cases, quantitative
investigation of the spectrum after telluric correction should be
conducted through proper evaluation of the smearing effects or through
the use of other approaches such as deconvolution of the observed
spectrum with the instrumental profile; this, however, is beyond the
scope of this paper.

%%%%%%%%%%%%%%%%%%%%%
% reference
%%%%%%%%%%%%%%%%%%%%%

%\bibliographystyle{apj}


\begin{thebibliography}{}

\bibitem[{{Beer}(1852)}]{1852AnP...162...78B}
{Beer}. 1852, Annalen der Physik, 162, 78

\bibitem[{{Bertaux} {et~al.}(2014){Bertaux}, {Lallement}, {Ferron}, {Boonne},
  \& {Bodichon}}]{2014A&A...564A..46B}
{Bertaux}, J.~L., {Lallement}, R., {Ferron}, S., {Boonne}, C., \& {Bodichon},
  R. 2014, \aap, 564, A46

\bibitem[{{Clough} {et~al.}(2005){Clough}, {Shephard}, {Mlawer}, {Delamere},
  {Iacono}, {Cady-Pereira}, {Boukabara}, \& {Brown}}]{2005JQSRT..91..233C}
{Clough}, S.~A., {Shephard}, M.~W., {Mlawer}, E.~J., {et~al.} 2005, \jqsrt, 91,
  233

\bibitem[{{Cotton} {et~al.}(2014){Cotton}, {Bailey}, \&
  {Kedziora-Chudczer}}]{2014MNRAS.439..387C}
{Cotton}, D.~V., {Bailey}, J., \& {Kedziora-Chudczer}, L. 2014, \mnras, 439,
  387

\bibitem[{{Gullikson} {et~al.}(2014){Gullikson}, {Dodson-Robinson}, \&
  {Kraus}}]{2014AJ....148...53G}
{Gullikson}, K., {Dodson-Robinson}, S., \& {Kraus}, A. 2014, \aj, 148, 53

\bibitem[{{Hamano} {et~al.}(2015){Hamano}, {Kobayashi}, {Kondo}, {Ikeda},
  {Nakanishi}, {Yasui}, {Mizumoto}, {Matsunaga}, {Fukue}, {Mito}, {Yamamoto},
  {Izumi}, {Nakaoka}, {Kawanishi}, {Kitano}, {Otsubo}, {Kinoshita},
  {Kobayashi}, \& {Kawakita}}]{2015ApJ...800..137H}
{Hamano}, S., {Kobayashi}, N., {Kondo}, S., {et~al.} 2015, \apj, 800, 137

\bibitem[{{Hamano} {et~al.}(2016){Hamano}, {Kobayashi}, {Kondo}, {Sameshima},
  {Nakanishi}, {Ikeda}, {Yasui}, {Mizumoto}, {Matsunaga}, {Fukue}, {Yamamoto},
  {Izumi}, {Mito}, {Nakaoka}, {Kawanishi}, {Kitano}, {Otsubo}, {Kinoshita}, \&
  {Kawakita}}]{2016ApJ...821...42H}
---. 2016, \apj, 821, 42

\bibitem[{{Hanson} {et~al.}(1996){Hanson}, {Conti}, \&
  {Rieke}}]{1996ApJS..107..281H}
{Hanson}, M.~M., {Conti}, P.~S., \& {Rieke}, M.~J. 1996, \apjs, 107, 281

\bibitem[{{Ikeda} {et~al.}(2016){Ikeda}, {Kobayashi}, {Kondo}, {Otsubo},
  {Hamano}, {Sameshima}, {Yoshikawa}, {Fukue}, {Nakanishi}, {Kawanishi},
  {Nakaoka}, {Kinoshita}, {Kitano}, {Asano}, {Takenaka}, {Watase}, {Mito},
  {Yasui}, {Minami}, {Izumu}, {Yamamoto}, {Mizumoto}, {Arasaki}, {Arai},
  {Matsunaga}, \& {Kawakita}}]{2016SPIE.9908E..5ZI}
{Ikeda}, Y., {Kobayashi}, N., {Kondo}, S., {et~al.} 2016, in \procspie, Vol.
  9908, Ground-based and Airborne Instrumentation for Astronomy VI, 99085Z

\bibitem[{{Kausch} {et~al.}(2015){Kausch}, {Noll}, {Smette}, {Kimeswenger},
  {Barden}, {Szyszka}, {Jones}, {Sana}, {Horst}, \&
  {Kerber}}]{2015A&A...576A..78K}
{Kausch}, W., {Noll}, S., {Smette}, A., {et~al.} 2015, \aap, 576, A78

\bibitem[{{Kurucz}(1993)}]{1993sssp.book.....K}
{Kurucz}, R.~L. 1993, {SYNTHE spectrum synthesis programs and line data}

\bibitem[{{Maiolino} {et~al.}(1996){Maiolino}, {Rieke}, \&
  {Rieke}}]{1996AJ....111..537M}
{Maiolino}, R., {Rieke}, G.~H., \& {Rieke}, M.~J. 1996, \aj, 111, 537

\bibitem[{{Markwardt}(2009)}]{2009ASPC..411..251M}
{Markwardt}, C.~B. 2009, in Astronomical Society of the Pacific Conference
  Series, Vol. 411, Astronomical Data Analysis Software and Systems XVIII, ed.
  D.~A. {Bohlender}, D.~{Durand}, \& P.~{Dowler}, 251

\bibitem[{{Prugniel} {et~al.}(2011){Prugniel}, {Vauglin}, \&
  {Koleva}}]{2011A&A...531A.165P}
{Prugniel}, P., {Vauglin}, I., \& {Koleva}, M. 2011, \aap, 531, A165

\bibitem[{{Rothman} {et~al.}(2009){Rothman}, {Gordon}, {Barbe}, {Benner},
  {Bernath}, {Birk}, {Boudon}, {Brown}, {Campargue}, {Champion}, {Chance},
  {Coudert}, {Dana}, {Devi}, {Fally}, {Flaud}, {Gamache}, {Goldman},
  {Jacquemart}, {Kleiner}, {Lacome}, {Lafferty}, {Mandin}, {Massie},
  {Mikhailenko}, {Miller}, {Moazzen-Ahmadi}, {Naumenko}, {Nikitin}, {Orphal},
  {Perevalov}, {Perrin}, {Predoi-Cross}, {Rinsland}, {Rotger}, {{\v S}ime{\v
  c}kov{\'a}}, {Smith}, {Sung}, {Tashkun}, {Tennyson}, {Toth}, {Vandaele}, \&
  {Vander Auwera}}]{2009JQSRT.110..533R}
{Rothman}, L.~S., {Gordon}, I.~E., {Barbe}, A., {et~al.} 2009, \jqsrt, 110, 533

\bibitem[{{Rudolf} {et~al.}(2016){Rudolf}, {G{\"u}nther}, {Schneider}, \&
  {Schmitt}}]{2016A&A...585A.113R}
{Rudolf}, N., {G{\"u}nther}, H.~M., {Schneider}, P.~C., \& {Schmitt},
  J.~H.~M.~M. 2016, \aap, 585, A113

\bibitem[{{Seifahrt} {et~al.}(2010){Seifahrt}, {K{\"a}ufl}, {Z{\"a}ngl},
  {Bean}, {Richter}, \& {Siebenmorgen}}]{2010A&A...524A..11S}
{Seifahrt}, A., {K{\"a}ufl}, H.~U., {Z{\"a}ngl}, G., {et~al.} 2010, \aap, 524,
  A11

\bibitem[{{Smette} {et~al.}(2015){Smette}, {Sana}, {Noll}, {Horst}, {Kausch},
  {Kimeswenger}, {Barden}, {Szyszka}, {Jones}, {Gallenne}, {Vinther},
  {Ballester}, \& {Taylor}}]{2015A&A...576A..77S}
{Smette}, A., {Sana}, H., {Noll}, S., {et~al.} 2015, \aap, 576, A77

\bibitem[{{Taniguchi} {et~al.}(2018){Taniguchi}, {Matsunaga}, {Kobayashi},
  {Fukue}, {Hamano}, {Ikeda}, {Kawakita}, {Kondo}, {Sameshima}, \&
  {Yasui}}]{2018MNRAS.473.4993T}
{Taniguchi}, D., {Matsunaga}, N., {Kobayashi}, N., {et~al.} 2018, \mnras, 473,
  4993

\bibitem[{{Vacca} {et~al.}(2003){Vacca}, {Cushing}, \&
  {Rayner}}]{2003PASP..115..389V}
{Vacca}, W.~D., {Cushing}, M.~C., \& {Rayner}, J.~T. 2003, \pasp, 115, 389

\end{thebibliography}
\end{document}